\documentclass[preprint,12pt]{elsarticle}

\usepackage{graphicx}
\usepackage{amsmath,amssymb,amsfonts}

\usepackage[mathlines]{lineno}
\usepackage{array}
\usepackage{subfigure}

\journal{Elsevier}

\begin{document}

\begin{frontmatter}


\title{Time-series learning of latent-space dynamics for reduced-order model closure}


\author[label1]{Romit Maulik\corref{cor1}}
\cortext[cor1]{Corresponding author}
\ead{rmaulik@anl.gov}
\author[label2]{Arvind Mohan}
\ead{arvindm@lanl.gov}
\author[label1]{Bethany Lusch}
\ead{blusch@anl.gov}
\author[label3]{Sandeep Madireddy}
\ead{smadireddy@anl.gov}
\author[label1,label3]{Prasanna Balaprakash}
\ead{pbalapra@anl.gov}
\author[label2]{Daniel Livescu}
\ead{livescu@lanl.gov}

\address[label1]{Argonne Leadership Computing Facility, Argonne National Laboratory, Lemont, IL 60439, USA\fnref{label1}}
\address[label2]{Center for Nonlinear Studies/CCS-2 Division, Los Alamos National Laboratory, Los Alamos, NM 87545, USA\fnref{label2}}
\address[label3]{Mathematics and Computer Science Division, Argonne National Laboratory, Lemont, IL 60439, USA\fnref{label3}}


\begin{abstract}
We study the performance of long short-term memory networks (LSTMs) and neural ordinary differential equations (NODEs) in learning latent-space representations of dynamical equations for an advection-dominated problem given by the viscous Burgers equation. Our formulation is devised in a non-intrusive manner with an equation-free evolution of dynamics in a reduced space with the latter being obtained through a proper orthogonal decomposition. In addition, we leverage the sequential nature of learning for both LSTMs and NODEs to demonstrate their capability for closure in systems which are not completely resolved in the reduced space. We assess our hypothesis for two advection-dominated problems given by the viscous Burgers equation. It is observed that both LSTMs and NODEs are able to reproduce the effects of the absent scales for our test cases more effectively than intrusive dynamics evolution through a Galerkin projection. This result empirically suggests that time-series learning techniques implicitly leverage a memory kernel for coarse-grained system closure as is suggested through the Mori-Zwanzig formalism.
\end{abstract}


\end{frontmatter}


\section{Introduction}
\label{Intro}

High-fidelity simulations of systems characterized by nonlinear partial differential equations represent immense computational expenditure and are prohibitive for decision-making tasks for applications. To address this issue, there has recently been a significant quantity of research into the reduced-order modeling (ROM) of such systems to reduce the degrees of freedom of the forward problem to manageable magnitudes \cite{carlberg2011efficient,wang2012proper,san2015principal,ballarin2015supremizer,san2018extreme,wang2019non,choi2019space}. As such, this field finds extensive application in control \cite{proctor2016dynamic}, multi-fidelity optimization \cite{peherstorfer2016optimal} and uncertainty quantification \cite{sapsis2013statistically,zahr2018efficient} among others. However, ROMs are limited in how they handle nonlinear dependence and perform poorly for complex physical phenomena which are inherently multiscale in space and time \cite{wells2017evolve,xie2018data,san2018neural,san2019artificial}. To address this issue, researchers continue to search for efficient and reliable ROM techniques for transient nonlinear systems. 

A common ROM development procedure may be described by the following tasks:
\begin{enumerate}
    \item Reduced basis identification.
    \item Nonlinear dynamical system evolution in the reduced basis.
    \item Reconstruction in full-order space for assessments.
\end{enumerate}

The first two items of the aforementioned schema individually constitute areas of extensive investigation, and there are studies which attempt to combine these into one optimization problem as well. In this investigation, we utilize conventional ideas for reduced basis identification with the use of the proper orthogonal decomposition (POD) for finding the optimal global basis. We proceed by considering a parameterized time-dependent partial differential equation given (in the full-order space) by
\begin{linenomath*}
\begin{align}
\label{gen1}
\dot{u}(x,t,\nu) + \mathcal{N}[u(x,t,\nu)] + \mathcal{L}[u(x,t,\nu); \nu] = 0, \quad (x,t,\nu) \in \Omega \times \mathcal{T} \times \mathcal{P},
\end{align}
\end{linenomath*}
where $\Omega \subset \mathbb{R}^1, \mathcal{T} = [0,T], \mathcal{P} \subset \mathbb{R}^1$ and $\mathcal{N}$, $\mathcal{L}$ are non-linear and linear operators respectively. Our system is characterized by a solution field $u : \Omega \times \mathcal{T} \times \mathcal{P} \rightarrow \mathbb{R}^1$ and appropriately chosen initial as well as boundary conditions.  We assume that our system of equations can be solved in space-time on a discrete grid resulting in the following systems of parameterized ODEs
\begin{linenomath*}
\begin{align}
\dot{\mathbf{u}_h}(t,\nu) + \mathbf{N}_{h}[\mathbf{u}_h(t,\nu)] + \mathbf{L}_h[\mathbf{u}_h(t,\nu); \nu] = 0  \quad(t,\nu) \in \mathcal{T} \times \mathcal{P},
\end{align}
\end{linenomath*}
where $\mathbf{u}_h : \mathcal{T} \times \mathcal{P} \rightarrow \mathbb{R}^{N_h}$ is a discrete solution and $N_h$ is the number of spatial degrees of freedom. Specifically, our problem is given by the viscous Burgers' equation with periodic boundary conditions which can be represented as
\begin{linenomath*}
\begin{align}
\begin{gathered}
\label{gen3}
\dot{u} + u\frac{\partial u}{\partial x} = \nu \frac{\partial^2 u}{\partial x^2}, \\
u(x,0) = u_0, \quad x \in [0,L], \quad u(0,t) = u(L,t) = 0.
\end{gathered}
\end{align}
\end{linenomath*}
It is well known that the above equations are capable of generating discontinuous solutions even if initial conditions are smooth and $\nu$ is sufficiently small due to advection-dominated behavior. We can then proceed to project our governing equations onto a space of reduced orthonormal bases for inexpensive forward solves of the dynamics. 

\subsection{Proper orthogonal decomposition}

In this section, we review the proper orthogonal decomposition (POD) technique for the construction of a reduced basis \cite{kosambi1943statistics,berkooz1993proper}. The interested reader may also find an excellent explanation of POD and its relationship with other dimension-reduction techniques in \cite{taira2019modal}. The POD procedure is tasked with identifying a space
\begin{linenomath*}
\begin{align}
\mathbf{X}^{f}=\operatorname{span}\left\{\boldsymbol{\vartheta}^{1}, \dots, \boldsymbol{\vartheta}^{f}\right\},
\end{align}
\end{linenomath*}
which approximates snapshots optimally with respect to the $L^2-$norm. The process of $\boldsymbol{\vartheta}$ generation commences with the collections of snapshots in the \emph{snapshot matrix}
\begin{linenomath*}
\begin{align}
\mathbf{S} = [\begin{array}{c|c|c|c}{\hat{\mathbf{u}}^{1}_h} & {\hat{\mathbf{u}}^{2}_h} & {\cdots} & {\hat{\mathbf{u}}^{N_{s}}_h}\end{array}] \in \mathbb{R}^{N_{h} \times N_{s}},
\end{align}
\end{linenomath*}
where $\hat{\mathbf{u}}_i : \mathcal{T} \times \mathcal{P} \rightarrow \mathbb{R}^{N_h}$ corresponds to an individual snapshot in time (for a total of $N_s$ snapshots) of the discrete solution domain with mean value removed i.e.,
\begin{linenomath*}
\begin{align}
\begin{gathered}
\hat{\mathbf{u}}^i_h = \mathbf{u}^i_h - \mathbf{\bar{u}}_h, \\
\mathbf{\bar{u}}_h = \frac{1}{N_s} \sum_{i=1}^{N_s} \mathbf{u}^i_h.
\end{gathered}
\end{align}
\end{linenomath*}
with $\overline{\mathbf{u}}_i : \mathcal{P} \rightarrow \mathbb{R}^{N_h}$ being the time averaged solution field. Our POD bases can then be extracted efficiently through the method of snapshots where we solve an eigenvalue problem for a correlation matrix
\begin{linenomath*}
\begin{align}
\begin{gathered}
\mathbf{C} \mathbf{W} = \Lambda \mathbf{W}, \\
\mathbf{C} = \mathbf{S}^T \mathbf{S} \in \mathbb{R}^{N_s \times N_s},
\end{gathered}
\end{align}
\end{linenomath*}
where $\Lambda = \operatorname{diag}\left\{\lambda_{1}, \lambda_{2}, \cdots, \lambda_{N_{s}}\right\} \in \mathbb{R}^{N_{s} \times N_{s}}$ is the diagonal matrix of eigenvalues and $\mathbf{W} \in \mathbb{R}^{N_{s} \times N_{s}}$ is the eigenvector matrix. Our POD basis matrix can then be obtained by
\begin{linenomath*}
\begin{align}
\begin{gathered}
\boldsymbol{\vartheta} = \mathbf{S} \mathbf{W} \in \mathbb{R}^{N_h \times N_s}.
\end{gathered}
\end{align}
\end{linenomath*}
In practice a reduced basis $\boldsymbol{\psi} \in \mathbb{R}^{N_h \times N_r}$ is built by choosing the first $N_r$ columns of $\boldsymbol{\vartheta}$ for the purpose of efficient ROMs where $N_r \ll N_s$. This reduced basis spans a space given by
\begin{linenomath*}
\begin{align}
\mathbf{X}^{r}=\operatorname{span}\left\{\boldsymbol{\psi}^{1}, \dots, \boldsymbol{\psi}^{N_r}\right\}.
\end{align}
\end{linenomath*}
The coefficients of this reduced basis (which capture the underlying temporal effects) may be extracted as
\begin{linenomath*}
\begin{align}
\begin{gathered}
\mathbf{A} = \boldsymbol{\psi}^{T} \mathbf{S} \in \mathbb{R}^{N_r \times N_s}.
\end{gathered}
\end{align}
\end{linenomath*}
The POD approximation of our solution is then obtained via
\begin{linenomath*}
\begin{align}
\hat{\mathbf{S}} =  [\begin{array}{c|c|c|c}{\tilde{\mathbf{u}}^{1}_h} & {\tilde{\mathbf{u}}^{2}_h} & {\cdots} & {\tilde{\mathbf{u}}^{N_{s}}_h}\end{array}] \approx \boldsymbol{\psi} \mathbf{A} \in \mathbb{R}^{N_h \times N_s},
\end{align}
\end{linenomath*}
where $\tilde{\mathbf{u}}_h^i : \mathcal{T} \times \mathcal{P} \rightarrow \mathbb{R}^{N_h}$ corresponds to the POD approximation to $\hat{\mathbf{u}}_h^i$. The optimal nature of reconstruction may be understood by defining the relative projection error
\begin{linenomath*}
\begin{align}
\frac{\sum_{i=1}^{N_{s}}\left\|\hat{\mathbf{u}}^i_h-\tilde{\mathbf{u}}^i_h \right\|_{\mathbb{R}^{N_{h}}}^{2}}{\sum_{i=1}^{N_{s}}\left\|\hat{\mathbf{u}}^i_h\right\|_{\mathbb{R}^{N_{h}}}^{2}}=\frac{\sum_{i=N_r+1}^{N_{s}} \lambda_{i}^{2}}{\sum_{i=1}^{N_{s}} \lambda_{i}^{2}},
\end{align}
\end{linenomath*}
which exhibits that with increasing retention of POD bases, increasing reconstruction accuracy may be obtained. As shall be explained later, the coefficient matrix $\mathbf{A}$ forms our training data for time-series learning.

\subsection{Galerkin-projection onto reduced space}

The orthogonal nature of the POD basis may be leveraged for a Galerkin projection onto the reduced basis. We start by revisiting Equation (\ref{gen1}) written in the form of an evolution equation for fluctuation components i.e.,
\begin{linenomath*}
\begin{align}
\dot{\hat{\mathbf{u}}}_h(x,t,\nu) + \mathcal{N}_h[\hat{\mathbf{u}}_h(x,t,\nu)] + \mathcal{L}_h[\hat{\mathbf{u}}_h(x,t,\nu); \nu] = 0,
\end{align}
\end{linenomath*}
which can expressed in the reduced basis as 
\begin{linenomath*}
\begin{align}
\boldsymbol{\psi} \dot{\mathbf{a}_r}(t,\nu) + \mathcal{N}_h[\boldsymbol{\psi} \mathbf{a}_r(t,\nu)] + \mathcal{L}_h[\boldsymbol{\psi} \mathbf{a}_r(t,\nu); \nu] = 0,
\end{align}
\end{linenomath*}
where $\mathbf{a}_r : \mathcal{T} \times \mathcal{P} \rightarrow \mathbb{R}^{N_r}, \mathbf{a}_r  \in \alpha$ corresponds to the temporal coefficients at one time instant of the system evolution (i.e., equivalent to a particular column of $\mathbf{A}$). The orthogonal nature of the reduced basis can be leveraged to obtain
\begin{linenomath*}
\begin{align}
\dot{\mathbf{a}_r}(t,\nu) + \mathcal{N}_r[\mathbf{a}_r(t,\nu)] + \mathcal{L}_r[\mathbf{a}_r(t,\nu); \nu] = 0,
\end{align}
\end{linenomath*}
which we denote the POD Galerkin-projection formulation (POD-GP). Note that we have assumed that the residual generated by the truncated representation of the full-order model is orthogonal to the reduced basis. We note that it is precisely this assumption that necessitates closure. From the point of view of the Burgers' equations given in Equation (\ref{gen3}), our POD-GP implementation is given as 
\begin{linenomath*}
\begin{align}
\frac{d a_{k}}{d t}=b_{k}^{1}+b_{k}^{2}+\sum_{i=1}^{N_r}\left(L_{i k}^{1}+L_{i k}^{2}\right) a_{i}+\sum_{i=1}^{N_r} \sum_{j=1}^{N_r} N_{i j k} a_{i} a_{j}, \quad \text { for } \quad k=1,2, \ldots, N_r,
\end{align}
\end{linenomath*}
where $a_k : \mathcal{T} \times \mathcal{P} \rightarrow \mathbb{R}^{1}$ is one component of $\mathbf{a}_r$ and where
\begin{linenomath*}
\begin{align}
\begin{aligned} 
b_{k}^{1} &=\left(\nu L[\overline{\mathbf{u}}_h], \boldsymbol{\psi}^{k}\right), \\ 
b_{k}^{2} &=\left(N[\overline{\mathbf{u}}_h ; \overline{\mathbf{u}}_h], \boldsymbol{\psi}^{k}\right),  \\ 
L_{i k}^{1} &=\left(\nu L\left[\boldsymbol{\psi}^{i} \right], \boldsymbol{\psi}^{k} \right), \\ 
L_{i k}^{2} &=\left(N\left[\overline{\mathbf{u}}_h ; \boldsymbol{\psi}^{i}\right]+N\left[\boldsymbol{\psi}^{i} ; \overline{\mathbf{u}}_h\right], \boldsymbol{\psi}^{k}\right), \\ 
N_{i j k} &=\left(N\left[\boldsymbol{\psi}^{i} ; \boldsymbol{\psi}^{j}\right], \boldsymbol{\psi}^{k}\right),
\end{aligned}
\end{align}
\end{linenomath*}
are operators which can be computed offline (i.e, $b_k^1, L_{i k}^{1} : \mathcal{P} \rightarrow \mathbb{R}^{1}; b_k^2, L_{i k}^{2}, N_{i j k} \in \mathbb{R}^1$) and where we have defined an inner-product by
\begin{linenomath*}
\begin{align}
(f,g) = \int_\Omega fg d\Omega.
\end{align}
\end{linenomath*}
with $L[f] = \frac{\partial^2 f}{\partial x^2}$ and $N[f;g] = -f \frac{\partial g}{\partial x}$, the operators stemming from the Burgers' equation. It will be discussed and demonstrated later, that the absence of higher-basis nonlinear interactions causes errors in the forward evolution of this system of equations. Note that the POD-GP essentially consists of $N_r$ coupled ODEs and is solved by a standard total-variation diminishing third-order Runge-Kutta method. The reduced degrees of freedom lead to very efficient forward solves of the problem. Note that this transformed problem has initial conditions given by
\begin{linenomath*}
\begin{align}
\mathbf{a}_r(t=0)=\left(\boldsymbol{\psi}^T \hat{\mathbf{u}}_h(t=0) \right).
\end{align}
\end{linenomath*}

\subsection{Contribution}

In this article, we investigate strategies to bypass the POD-GP process with an intrusion-free (or equation-free) learning of dynamics in reduced space. We deploy machine learning strategies devised for sequential data learning on time-series samples extracted from the true dynamics of the problems considered. In recent literature, there has been a considerable interest in the utility of machine learning techniques for effective ROM construction. Data-driven techniques have been used in various phases of ROM techniques such as in reduced basis construction \cite{lee2018model}, augmented dynamics evolution \cite{lusch2018deep,mohan2018deep,san2018neural,san2018extreme,guo2019data,san2019artificial,mohebujjaman2019physically,wang2019non,yeo2019deep,mohan2019compressed} and system identification \cite{brunton2016discovering,kutz2016dynamic,rudy2017data,champion2019data,raissi2017physics}. 

In this article, we study the utility of data-driven techniques to make \emph{a posteriori} predictions for state evolution in reduced space with assessments made on their ability to reconstruct transient characteristics of the full-order solution. It is also observed that the ability to learn a time series (possible through the in-built design of memory and an assumption of non-i.i.d sequential data in the learning) leads to an implicit closure whereby the effects of uncaptured frequencies are retained, drawing parallels to a Mori-Zwanzig formalism. In related work, the study in \cite{ma2018model} utilizes recurrent neural networks to explicitly specify a sub-grid stress model for large eddy simulation with the lower frequency evolution controlled by coarse-grained partial differential equations. Our framework is also similar to \citep{wang2019recurrent} where an LSTM is utilized to learn a \emph{parametric} memory kernel for explicit closure of nonlinear PDEs. The present study can be considered a \emph{non-intrusive} counterpart of these investigations. In addition, we also detail a formalism for efficient machine learning architecture selection using scaleable Bayesian Optimization. Our test problems are given by the advection-dominated viscous Burgers equation \cite{san2018neural} with a moving shock as well as a pseudo-turbulence test case denoted `Burgulence' \cite{bec2007burgers,maulik2018explicit} showing the characteristic $k^{-2}$ scaling in wavenumber ($k$) space.

\section{Latent-space learning}

In this section, we outline our machine learning techniques for latent-space time-series learning. We study two techniques built around the premise of preserving memory effects within their architecture - neural ordinary differential equations (NODE) and long short-term memory networks (LSTMs). Both frameworks are tasked with predicting the evolution of $\mathbf{a}_r$ over time. 

\subsection{Neural ordinary differential equations}

In recent times, there have been several studies which have interpreted residual neural networks using dynamical systems theory \cite{haber2017stable,ruthotto2018deep,behrmann2018invertible,reshniak2019robust}. The framework of the NODE \cite{chen2018neural} envisions the learning of $\mathbf{a}_r$ over time as
\begin{linenomath*}
\begin{align}
\frac{d \mathbf{a}_r}{d t}=f(\mathbf{a}_r, \theta), \quad (\theta) \in \Theta,
\end{align}
\end{linenomath*}
where $\Theta \subset \mathbb{R}^{N_w}$ is a space of $N_w$ user-defined model parameters. The learning can be thought to be through a continuous backpropagation through time, i.e., where there an infinite number of layers in the network with the input layer given by $\mathbf{a}_r(t=0)$ and the output layer given by $\mathbf{a}_r(t=T)$. Therefore the NODE approximates the latent-space evolution as an ordinary differential equation which is continuous through time in a manner similar to the Galerkin projection. The function $f : \alpha \times \Theta \rightarrow \mathbb{R}^{N_r}$ in this study is represented by a neural network with a single 40-neuron hidden layer and a tan-sigmoid activation where $\alpha \subset \mathbb{R}^{N_r}, \Theta \subset \mathbb{R}^{N_w}$ and $N_w$ is the number of parameters of the neural network architecture. Note that the assumption of a single hidden layer architecture for the right-hand side of the latent-space ODE allows for upper-bound guarantees given by the universal approximation theorem (Barron, 1993) although more complicated dynamics may require deeper architectures. Readers are referred to the work of Chen et al. \cite{chen2018neural}, for a detailed discussion of the neural ODE and its utility in learning sequential data. 

The forward propagation of information through time (i.e., from $t=0$ to $t=T$) is performed through a standard ODE solver (in this case a first-order accurate Euler method) whereas backpropagation of errors is performed through the backward solve of an adjoint system given by
\begin{linenomath*}
\begin{align}
\label{adj}
\frac{d \mathbf{b}_g}{d t}=-\mathbf{b}_g^{\mathrm{T}} \frac{\partial f(\mathbf{a}_r, \theta)}{\partial \mathbf{a}_r},
\end{align}
\end{linenomath*}
where $\mathbf{b}_g : \mathcal{T} \times \alpha \times \mathcal{E} \rightarrow \mathbb{R}^{N_r + N_w + 1}$ is the augmented state vector given by 
\begin{linenomath*}
\begin{align}
\mathbf{b}_g = [\frac{\partial E}{\partial \mathbf{a}_r}, \frac{\partial E}{\partial \theta}, \frac{\partial E}{\partial t}]^T, \quad E \in \mathcal{E},
\end{align}
\end{linenomath*}
with scalar loss at final time $\mathcal{E} \subset \mathbb{R}^1$ obtained at the $t=T$ following forward-propagation. Each calculation of $E$ is followed by the backward solve of Equation (\ref{adj}) (which may be interpreted as continuous backpropagation in time) to calculate $\mathbf{b}_g (t=0)$ which can then be used to determine $\frac{\partial E}{\partial \theta} (t=0)$. This value of the gradient can then be used to update the parameters $\theta$ using an optimization algorithm. In this article, we utilize RMSProp for our loss minimization with a learning rate of 0.01 and a momentum parameter of 0.9. Instead of performing the forward deployment of the NODE and backpropagation of the model errors for the entire domain, we utilize 1000 samples of our total data as our training and validation dataset using the technique detailed in the original article to speed up training. Each sample is given by a sequence of 10 timesteps. The training, for each epoch, is performed using 10 randomly chosen samples (i.e., our batch size is 10) for the calculation of parameter gradients. The final gradient deployed for model training is averaged across this batch. A set of samples (20\% of the total 1000), chosen randomly before training, is kept aside from the learning process to assess validation errors. Note that validation errors are also characterized by final timestep loss (i.e., at timestep 10 of each batch) thereby incorporating the degree of error accummulation due to an inaccurately trained model at that epoch. The best model corresponds to the lowest validation loss (averaged across all validation samples). We do not utilize a separate dataset for the purpose of testing. 

For the purpose of testing, we note that all assessments for the problems are through forward (or \emph{a posteriori}) deployment. In other words, the NODE is specified an initial condition and then deployed to obtain state vectors using an ODE forward solve until the final time. The prediction at each time step is obtained by the Euler integration which requires the knowledge of previous state alone. Note that apart from the first prediction by NODE (which utilizes the initial condition), state predictions are recursively utilized for predicting the future. Therefore, testing may be assumed to be a long-term predictive test of the model learning in the presence of deployment error.

\subsection{Long short-term memory networks}

The long short-term memory (LSTM) network was introduced to consider time-delayed processes where events further back in the past may potentially affect predictions for the current location in the sequence. The basic equations of the LSTM in our context for an input variable $\mathbf{a}_r$ are given by
\begin{linenomath*}
\begin{align}
\begin{split}
\text{input gate: }& \boldsymbol{G}_{i}=\boldsymbol{\varphi}_{S} \circ \mathcal{F}_{i}^{N_{c}}(\mathbf{a}_r), \\
\text{forget gate: }& \boldsymbol{G}_{f}=\boldsymbol{\varphi}_{S} \circ \mathcal{F}_{f}^{N_{c}}(\mathbf{a}_r), \\
\text{output gate: }& \boldsymbol{G}_{o}=\boldsymbol{\varphi}_{S} \circ \mathcal{F}_{o}^{N_{c}}(\mathbf{a}_r), \\
\text{internal state: }& \boldsymbol{s}_{t}=\boldsymbol{G}_{f} \odot \boldsymbol{s}_{t-1}+\boldsymbol{G}_{i} \odot\left(\boldsymbol{\varphi}_{T} \circ \mathcal{F}_{\mathbf{a}_r}^{N_{c}}(\mathbf{a}_r)\right), \\
\text{output: }& \boldsymbol{G}_{o} \circ \boldsymbol{\varphi}_{T}\left(\boldsymbol{s}_{t}\right),
\end{split}
\end{align}
\end{linenomath*}
in which $\boldsymbol{\varphi}_{S}$ and $\boldsymbol{\varphi}_{L}$ refer to tangent sigmoid and tangent hyperbolic activation functions respectively, $N_c$ is the number of hidden layer units in the LSTM network. Note that $\mathcal{F}^{n}$ refers to a linear operation given by a matrix multiplication and subsequent bias addition i.e,
\begin{linenomath*}
\begin{align}
\mathcal{F}^{n}(\boldsymbol{x})=\boldsymbol{W} \boldsymbol{x}+\boldsymbol{B},
\end{align}
\end{linenomath*}
where $\boldsymbol{W} \in \mathbb{R}^{n \times m}$ and $\boldsymbol{B} \in \mathbb{R}^{n}$ for $\mathbf{x} \in \mathbb{R}^m$ and where $\mathbf{a} \circ \mathbf{b}$ refers to a Hadamard product of two vectors. The LSTM implementation will be used to advance $\mathbf{a}_r$ as a function of time in the reduced space. The LSTM network's primary utility is the ability to control information flow through time with the use of the gating mechanisms. A greater value of the forget gate (post sigmoidal activation), allows for a greater preservation of past state information through the sequential inference of the LSTM, whereas a smaller value suppresses the influence of the past. Our LSTM deployment utilized 32 neurons in its input, forget, output and state calculation operations each and utilized a learning rate of 0.001. It uses a sequence to sequence prediction utilized as a rolling window for predicting the output at the next timestep. We utilize a batch size of 16 samples with each sample having a sequence of 10 timesteps for all of our LSTM deployments. As in the previous learning approach, a set of data is kept aside for validation. This validation loss is used to make decisions about model selection. Note that the total number of samples (1000) is the same as the NODE deployment.

\subsection{Connection with Mori-Zwanzig formalism}

In this section, we outline the Mori-Zwanzig formalism \cite{mori1965transport,zwanzig1973nonlinear} for the viscous Burgers equation and connect it to time-series learning in POD space. We frame the (full-order) dynamics evolution in latent space using the following formulation derived from the first step of the Mori-Zwanzig treatment
\begin{linenomath*}
\begin{align}
\frac{d \textbf{a}_r}{d t} = e^{\mathcal{W} t} \mathcal{W} \mathbf{a}_{r},
\end{align}
\end{linenomath*}
where $\mathcal{W}$ is the viscous Burgers operator given by
\begin{linenomath*}
\begin{align}
\mathcal{W} = -\mathcal{N}_{r}\left[. \right]-\mathcal{L}_{r}\left[ . \right].
\end{align}
\end{linenomath*}

We proceed by defining two self-adjoint projection operators into orthogonal subspaces given by
\begin{linenomath*}
\begin{align}
P \textbf{a}_h = \frac{(\textbf{a}_r^0, \textbf{a}_h^0)}{(\textbf{a}_r^0,\textbf{a}_r^0)} \textbf{a}_r, \quad Q = I-P,
\end{align}
\end{linenomath*}
with $QP=0$ and $\textbf{a}^0 = \textbf{a} (t=0)$. Therefore, $P$ may be assumed to be a projection of our full-order representation in POD space ($\mathbf{a}_h$ living in $\textbf{X}^f$) onto the reduced basis ($\mathbf{a}_r$ living in $\mathbf{X}^r$). We can further expand our system as 
\begin{linenomath*}
\begin{align}
\frac{d \textbf{a}_r}{d t} = e^{\mathcal{W} t} (Q+P) \mathcal{W} \mathbf{a}_{r},
\end{align}
\end{linenomath*}
which may further be decoupled to a Markov-like projection operator $\mathcal{M}$ given by 
\begin{linenomath*}
\begin{align}
e^{\mathcal{W} t} P \mathcal{W} a_r =  \frac{(\textbf{a}_r^0, \textbf{a}_h^0)}{(\textbf{a}_r^0,\textbf{a}_r^0)} e^{\mathcal{W} t} a_r = \mathcal{M} a_r,
\end{align}
\end{linenomath*}
and a memory operator $\mathcal{G}$ given by
\begin{linenomath*}
\begin{align}
e^{\mathcal{W} t} Q \mathcal{W} \mathbf{a}_r^0 = e^{Q \mathcal{W} t} Q \mathcal{W} \mathbf{a}_r^0 + \int_{0}^{t} e^{\mathcal{W}\left(t-t_{1}\right)} P \mathcal{W} e^{Q \mathcal{W} t_{1}} Q \mathcal{W} \mathbf{a}_r^0 d t_{1} = \mathcal{G} a_r,
\end{align}
\end{linenomath*}
for which we have used Dyson's formula \cite{evans2008statistical} and where $t_1$ corresponds to a hyperparameter that specifies the length of memory retention. The second relationship may be assumed to be a combination of memory effects and noise. The final evolution of the system can then be bundled into a linear combination of these two kernels i.e.,
\begin{linenomath*}
\begin{align}
\frac{d \textbf{a}_r}{d t} = \mathcal{G} a_r + \mathcal{M} a_r.
\end{align}
\end{linenomath*}

The reader may compare this expression with that of the internal state update within an LSTM which we revisit here -
\begin{linenomath*}
\begin{align}
\boldsymbol{s}_{t}=\boldsymbol{G}_{f} \odot \boldsymbol{s}_{t-1}+\boldsymbol{G}_{i} \odot\left(\boldsymbol{\varphi}_{T} \circ \mathcal{F}_{\mathbf{a}_r}^{N_{c}}(\mathbf{a}_r)\right),
\end{align}
\end{linenomath*}
where a linear combination of a nonlinearly transformed input vector at time $t$ with the gated result of a hidden-state at a previous time $t-1$ is utilized for calculating the result vector at the current time. The process of carrying a state through time via gating may be assumed to be a representation of the memory integral (as well as the noise) whereas the utilization of the current input may be assumed to be the Markovian component of the map. In contrast, from the point of view of the NODE implementation, the goal is to learn $e^{\mathcal{W} t} \mathcal{W} \mathbf{a}_{r}$ directly through a neural network. 

\section{Experiments}

In this section, we assess the performance of both NODE and LSTM frameworks in representing latent-space dynamics appropriately. We investigate two problems given by the viscous Burgers' equation in a periodic domain. Both problems are advection dominated where the first has a moving discontinuity over time (which we shall designate the \emph{advecting shock} problem) and the second which is characterized by standing shocks of various magnitudes (which we designate \emph{Burgulence}). Their problem statement and results are shown below.

\subsection{Advecting shock}

\begin{figure}
	\centering
	\includegraphics[width=0.85\textwidth]{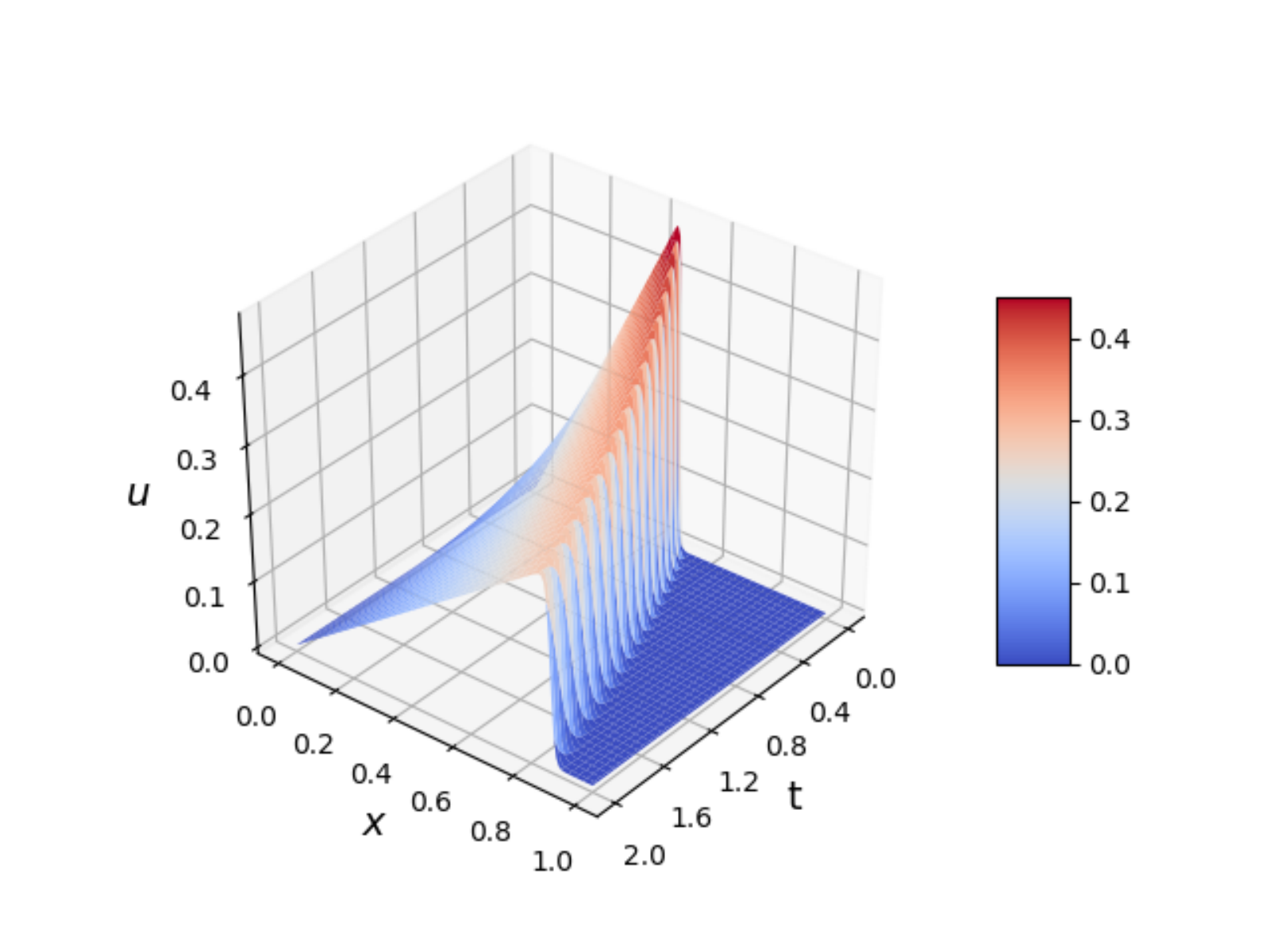}
	\caption{Full-order solution of the advecting shock problem. Note the presence of a moving discontinuity.}
	\label{Figure1}
\end{figure}

Our first problem is given by the following initial and boundary conditions
\begin{linenomath*}
\begin{align}
u(x, 0) &=\frac{x}{1+\sqrt{\frac{1}{t_{0}}} \exp \left(R e \frac{x^{2}}{4}\right)}, \\ 
u(0, t) &=0, \\ 
u(L, t) &=0,
\end{align}
\end{linenomath*}
where we specify $L=1$ and maximum time $t_{max}=2$. An analytical solution for the above set of equations exists and is given by
\begin{linenomath*}
\begin{align}
u(x, t)=\frac{\frac{x}{t+1}}{1+\sqrt{\frac{t+1}{t_{0}}} \exp \left(R e \frac{x^{2}}{4 t+4}\right)},
\end{align}
\end{linenomath*}
where $t_0=\text{exp}(Re/8)$ and $Re = 1/\nu$ is kept fixed at 1000. We directly utilize this expression to generate our snapshot data for ROM assessments. A visualization of the time evolution of the initial condition is shown in Figure \ref{Figure1}. As outlined in a previous assessment of this problem \cite{san2018neural}, a reduced basis constructed of 20 basis vectors retains 99.93\% of the energy of the system. For the purpose of our assessments, we retain solely three modes which results in an unresolved ROM which corresponds to only 86.71 \% of the total energy - thus necessitating closure. 

\begin{figure}
	\centering
	\subfigure[LSTM]{\includegraphics[width=0.48\textwidth]{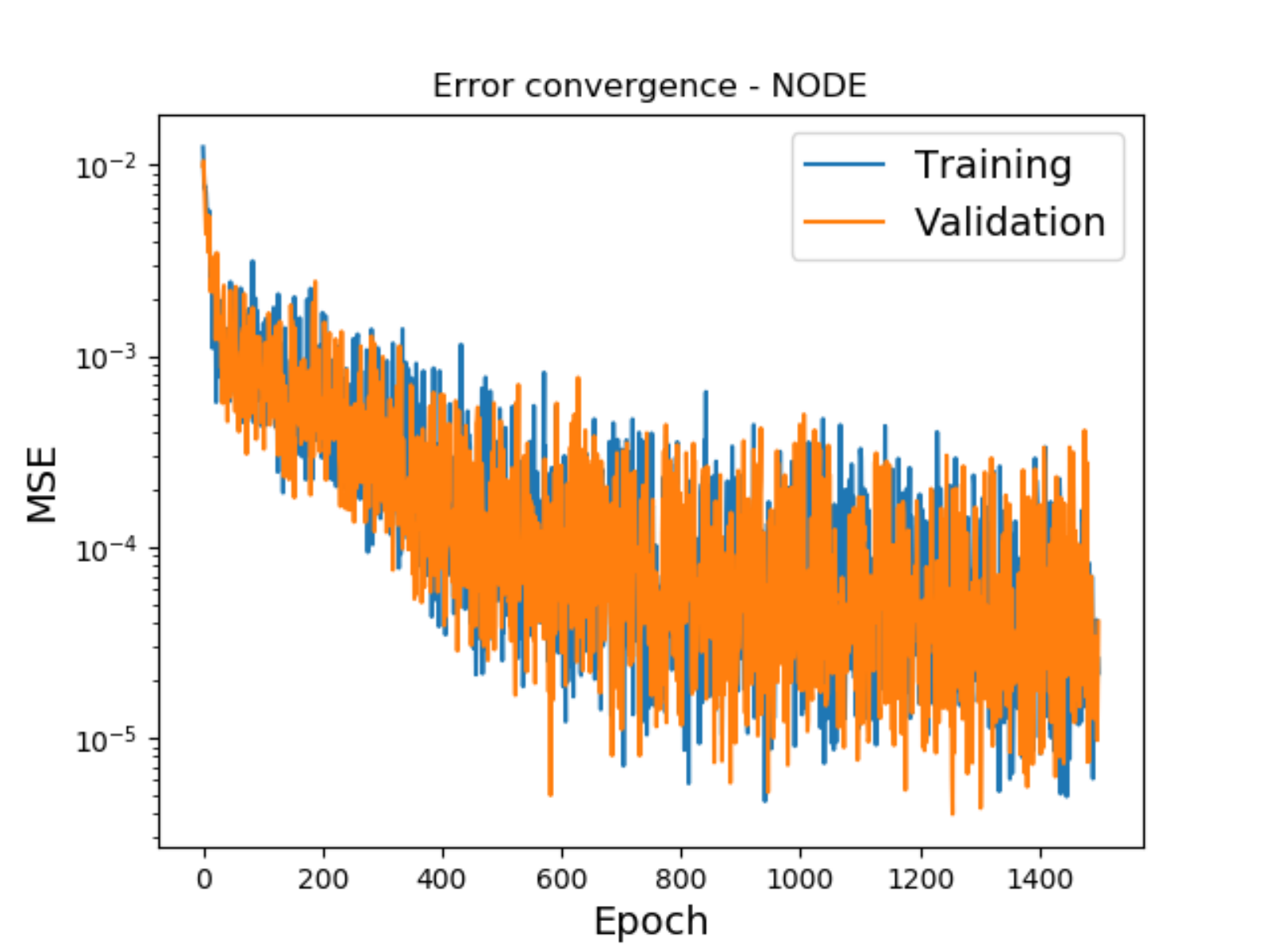}}
	\subfigure[NODE]{\includegraphics[width=0.48\textwidth]{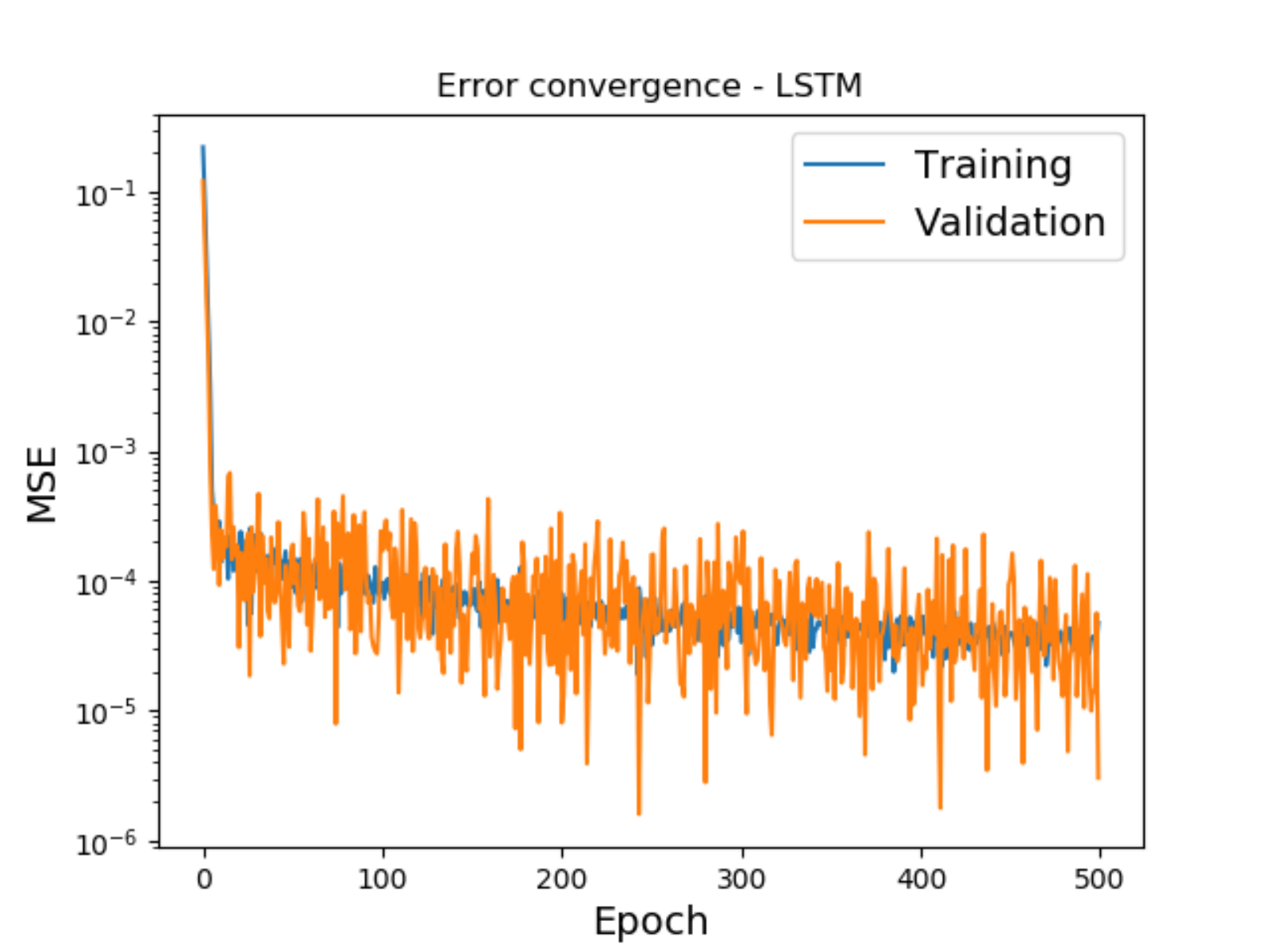}}
	\caption{Training and validation loss convergence with epochs for time-series predictions of the advecting shock case.}
	\label{Figure2}
\end{figure}

We perform an optimization for learning the modal coefficient time series using the LSTM and NODE frameworks. To recap model specifics, we deploy NODE (using 40 neurons) and LSTM (using 32 hidden layer neurons) for learning the sequential nature of the modal coefficient evolution in POD space. We utilize the RMSprop optimizer using a learning rate of 0.01 for the former and 0.001 for the latter and a momentum coefficient of 0.9 for both. Batch sizes of 10 and 16 respectively are also used at each epoch of the learning phase. 1000 randomly chosen sequence lengths of 10 are utilized for learning and validation through time with 20\% of the total data kept aside for the latter. We note that the best validation loss (aggregated over all validation samples) is utilized for model selection. Figure \ref{Figure2} shows the progress to convergence for both LSTM and NODE architectures during training for the first three modal coefficients. Both NODE and LSTM trainings are run until validation loss magnitudes hover around a magnitude of 10\textsuperscript{-4}. It is observed that the LSTM framework reaches convergence more quickly although the oscillating losses of the NODE potentially indicate better exploration. We note that the oscillations may also indicate the requirement of a lower learning rate.

\begin{figure}
	\centering
	\includegraphics[width=\textwidth]{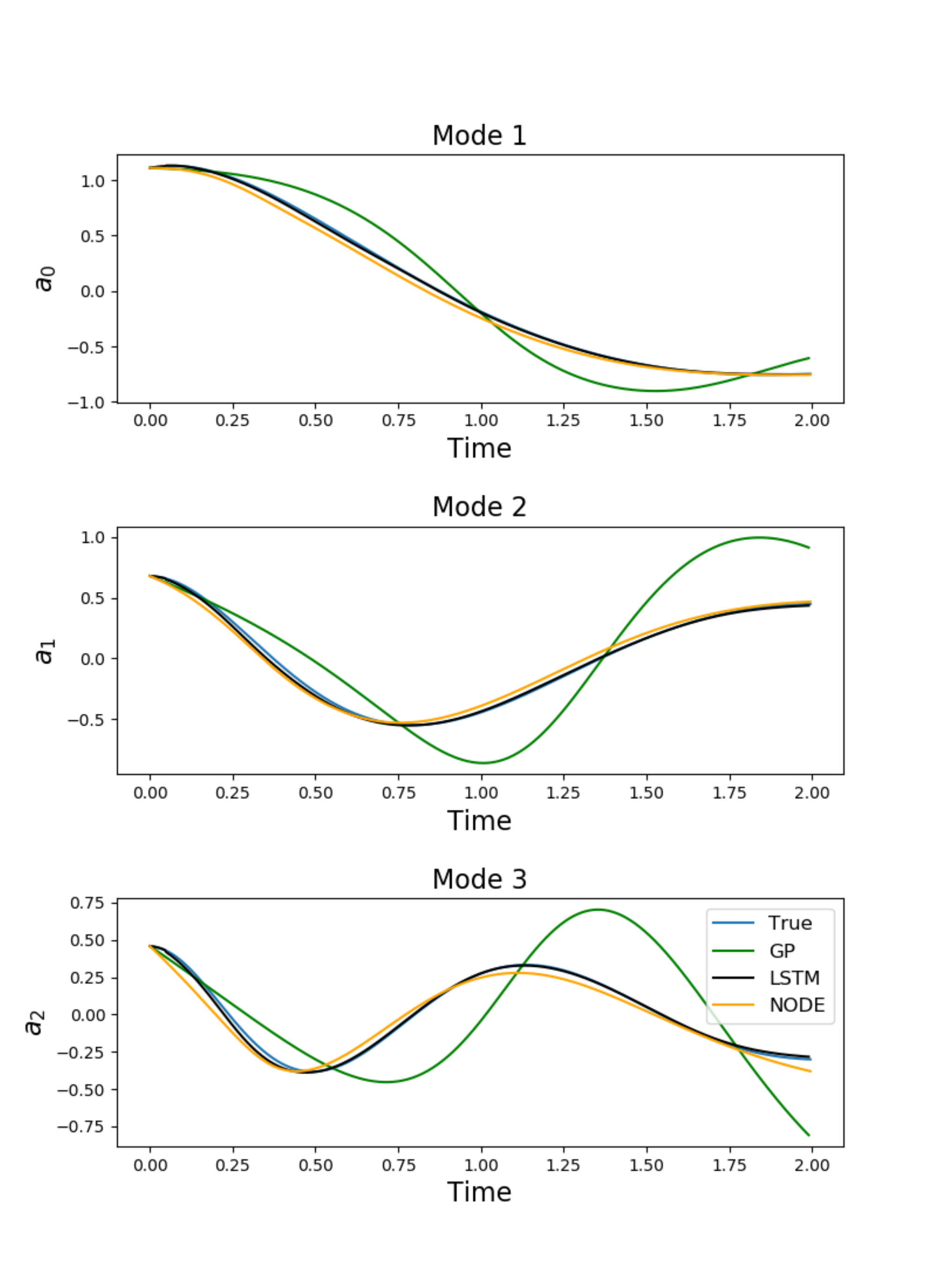}
	\caption{POD-space coefficient evolution for the advecting shock case.}
	\label{Figure3}
\end{figure}

The time-series predictions for the trained frameworks are shown in Figure \ref{Figure3} where $a_0, a_1$ and $a_2$ correspond to the first three retained modes. For the purpose of comparison, we also show predictions from GP and the true modal coefficients, the latter of which are utilized for training our time-series predictions. We can observe that both LSTM and NODE deployments capture coefficient trends accurately indicating that sequential behavior has been learned. The GP predictions can be seen to show unphysical growth in coefficient amplitude due to the lack of presence of the finer modes. However, LSTM and NODE deployments embed memory into their formulation in the form of a hidden state or through explicit learning of a latent-space ODE respectively. The memory-based nature of their learning leads to excellent agreement with the true behavior of the resolved scales.

\begin{figure}
	\centering
	\includegraphics[width=\textwidth]{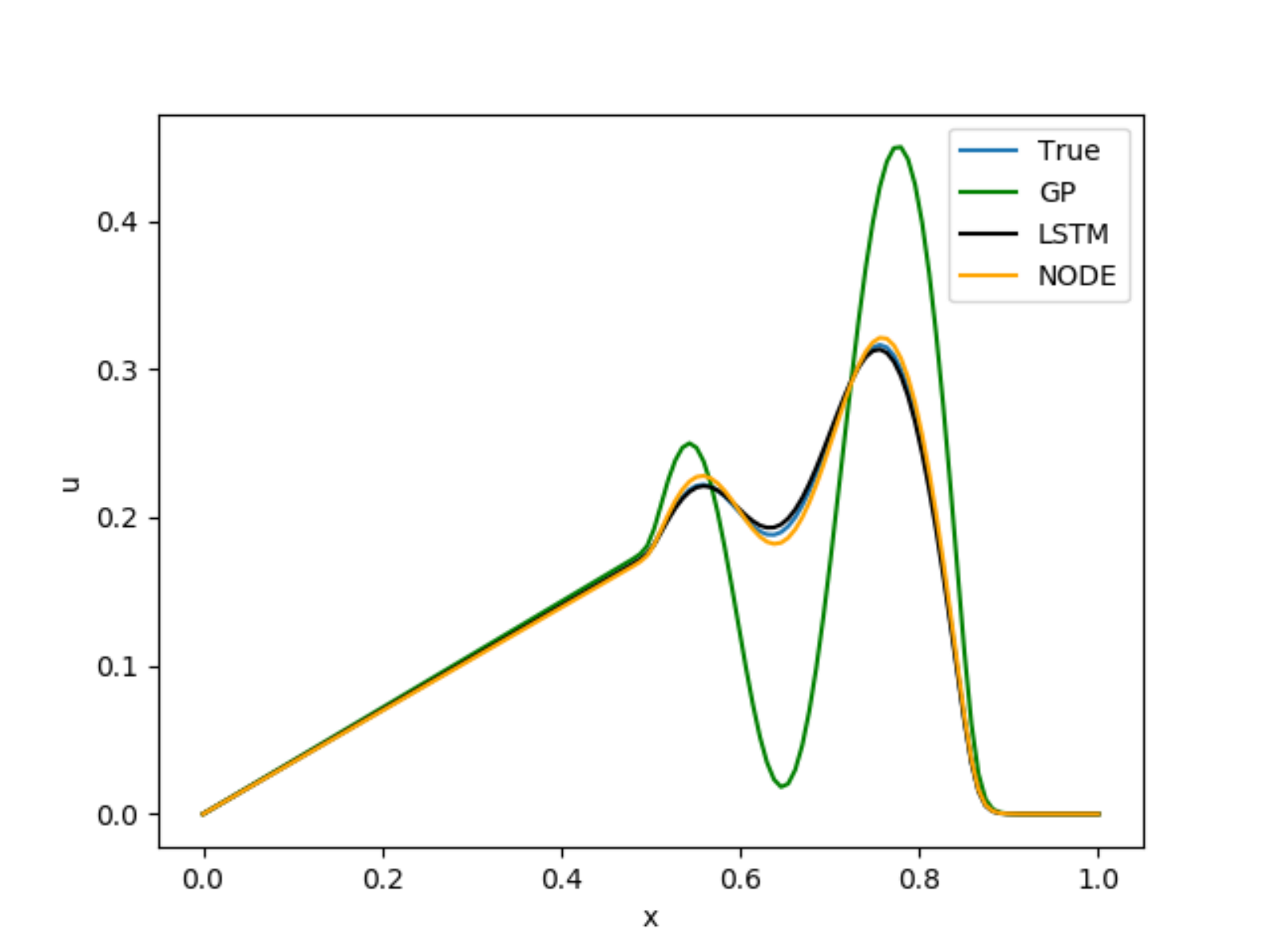}
	\caption{Field reconstruction ability for NODE and LSTM for the advecting shock case.}
	\label{Figure4}
\end{figure}

\begin{figure}
	\centering
	\subfigure[Field]{\includegraphics[width=0.48\textwidth]{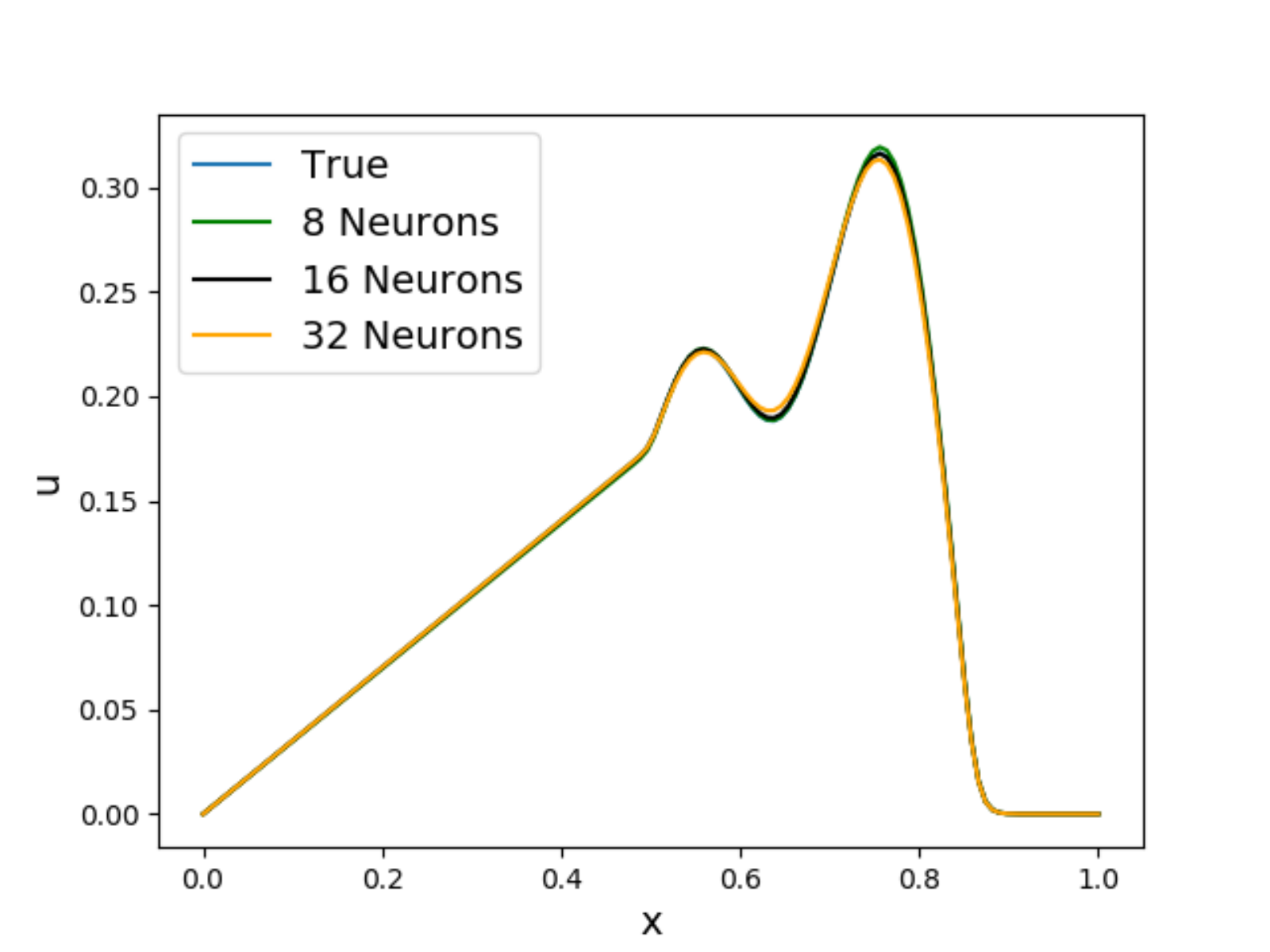}}
	\subfigure[Zoomed]{\includegraphics[width=0.48\textwidth]{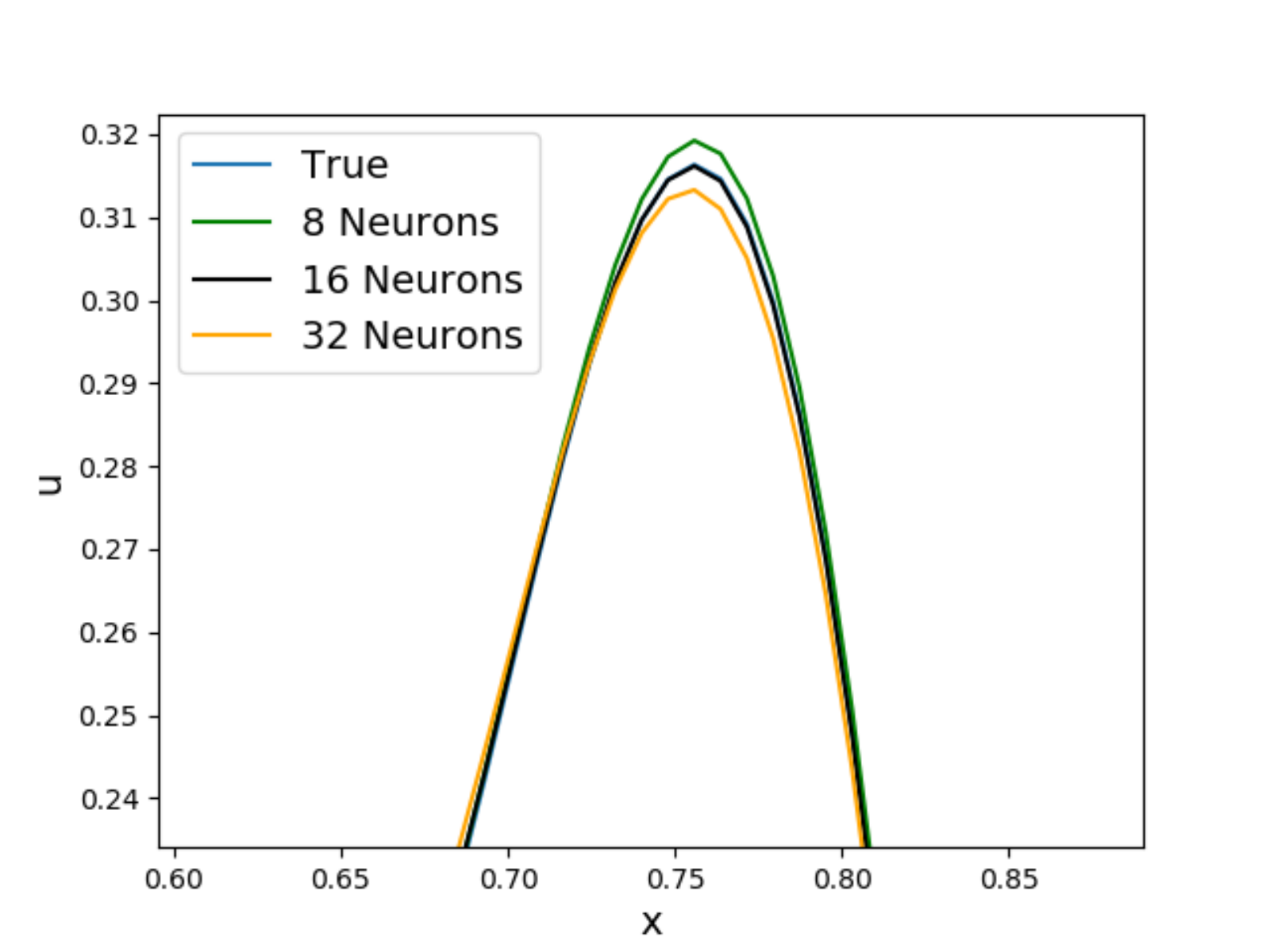}}
	\caption{A comparison of three different LSTM predictions for the advecting shock case.}
	\label{Figure5}
\end{figure}

\begin{figure}
	\centering
	\subfigure[Zoomed out]{\includegraphics[width=0.48\textwidth]{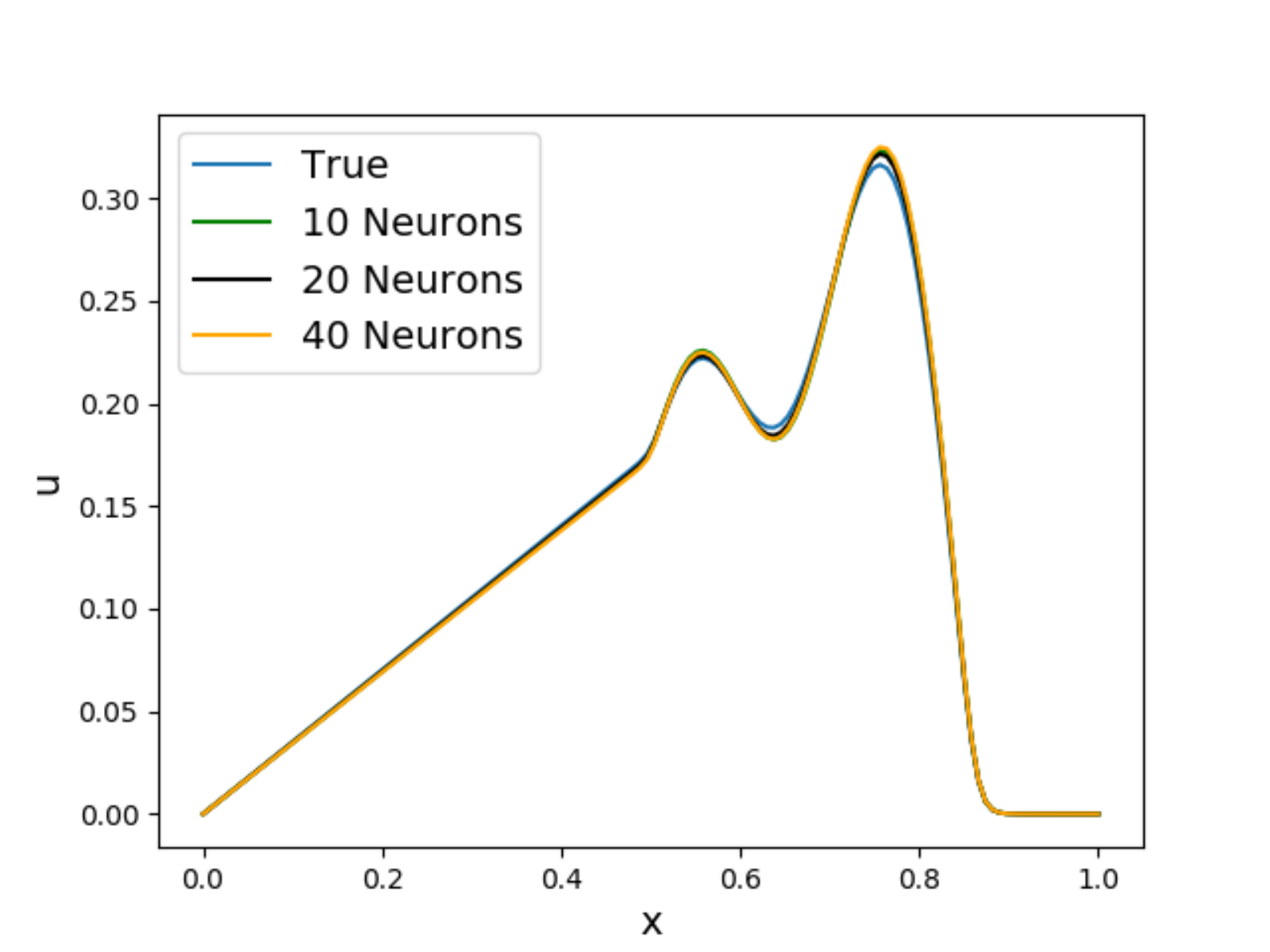}}
	\subfigure[Zoomed in]{\includegraphics[width=0.48\textwidth]{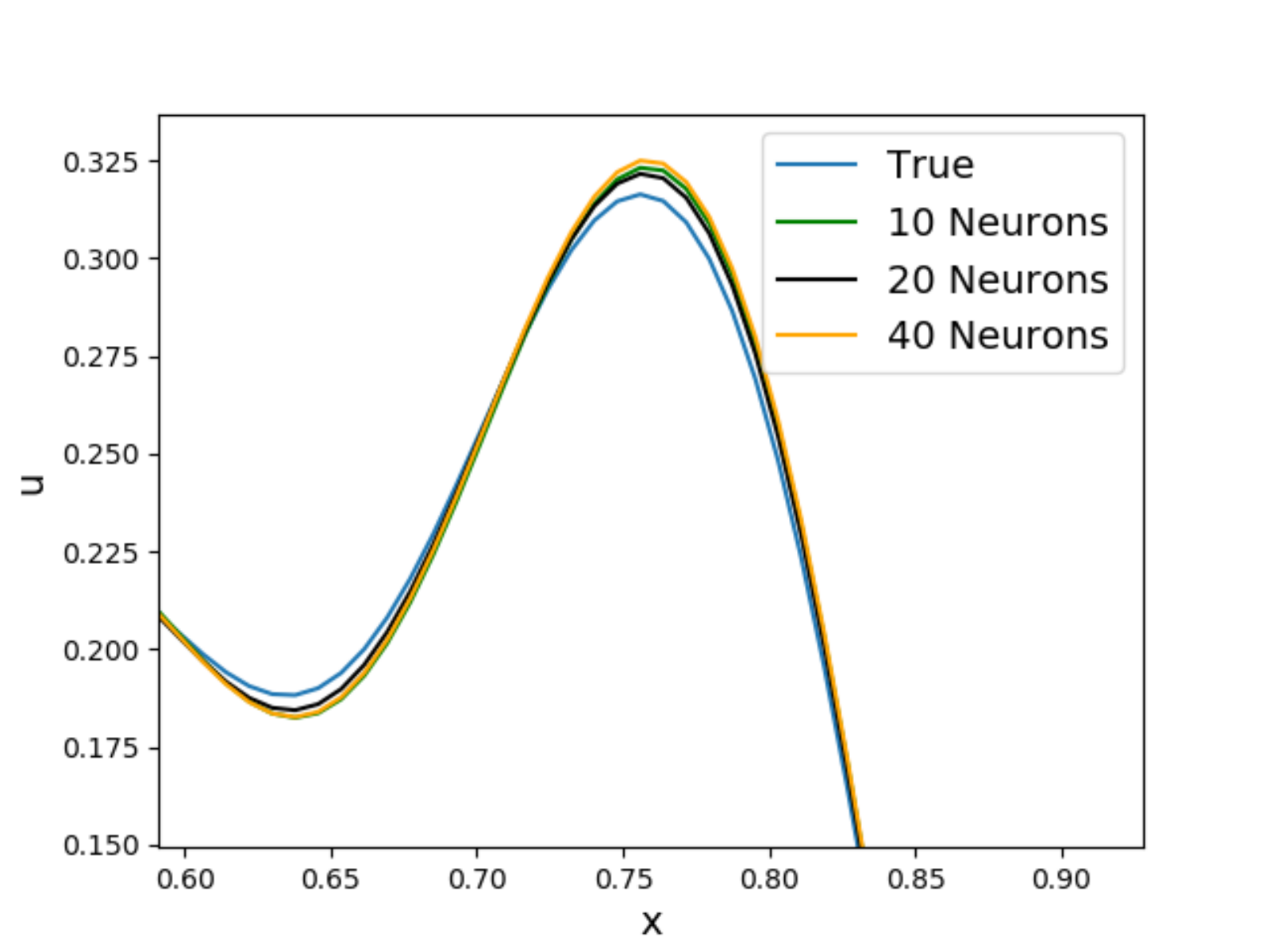}}
	\caption{A comparison of three different NODE predictions for the advecting shock case. The deployment with 16 neurons coincides with the true solution.}
	\label{Figure6}
\end{figure}

\begin{figure}
	\centering
	\includegraphics[width=\textwidth]{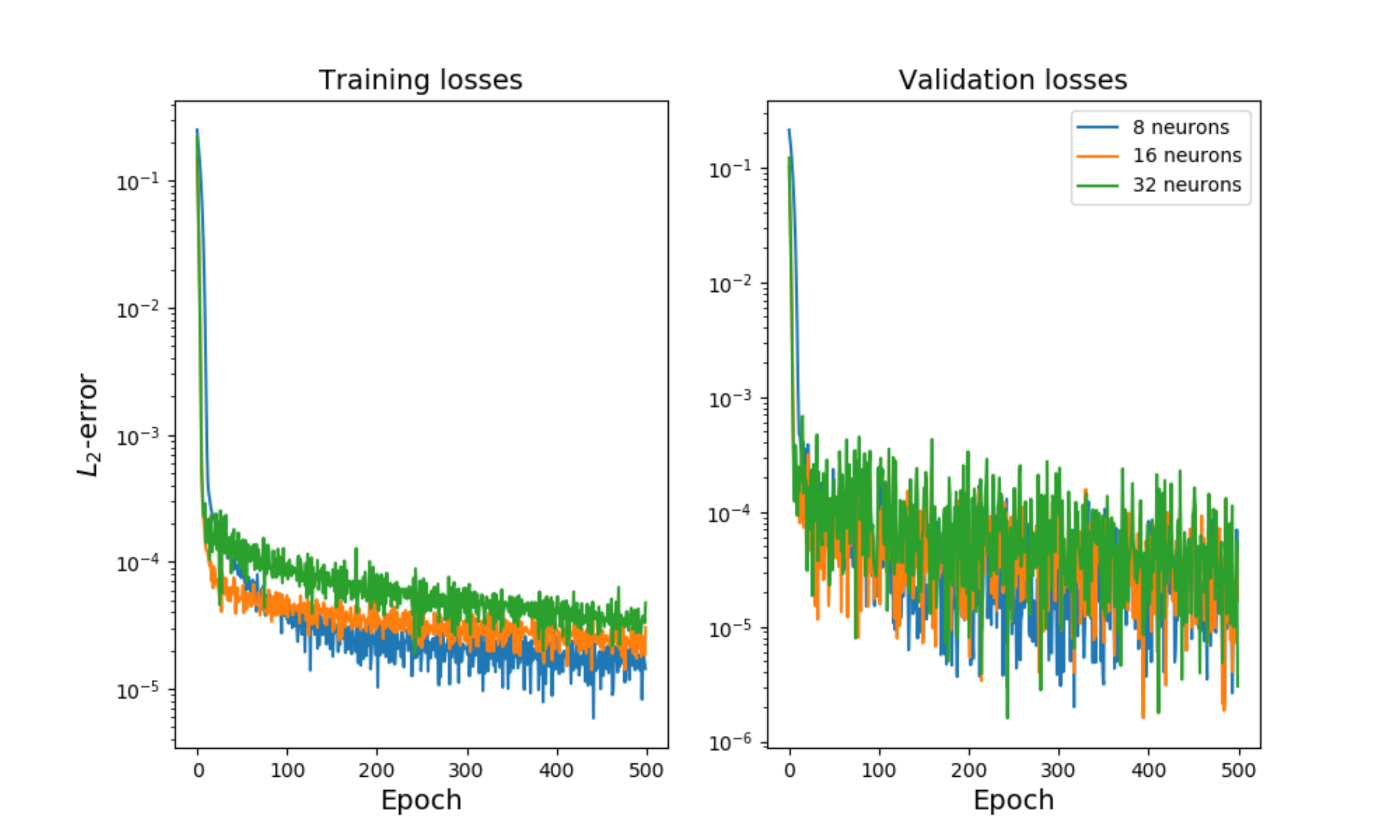}
	\caption{Ensemble training and validation losses for the LSTM architecture for the advecting shock case.}
	\label{Figure7}
\end{figure}

\begin{figure}
	\centering
	\includegraphics[width=\textwidth]{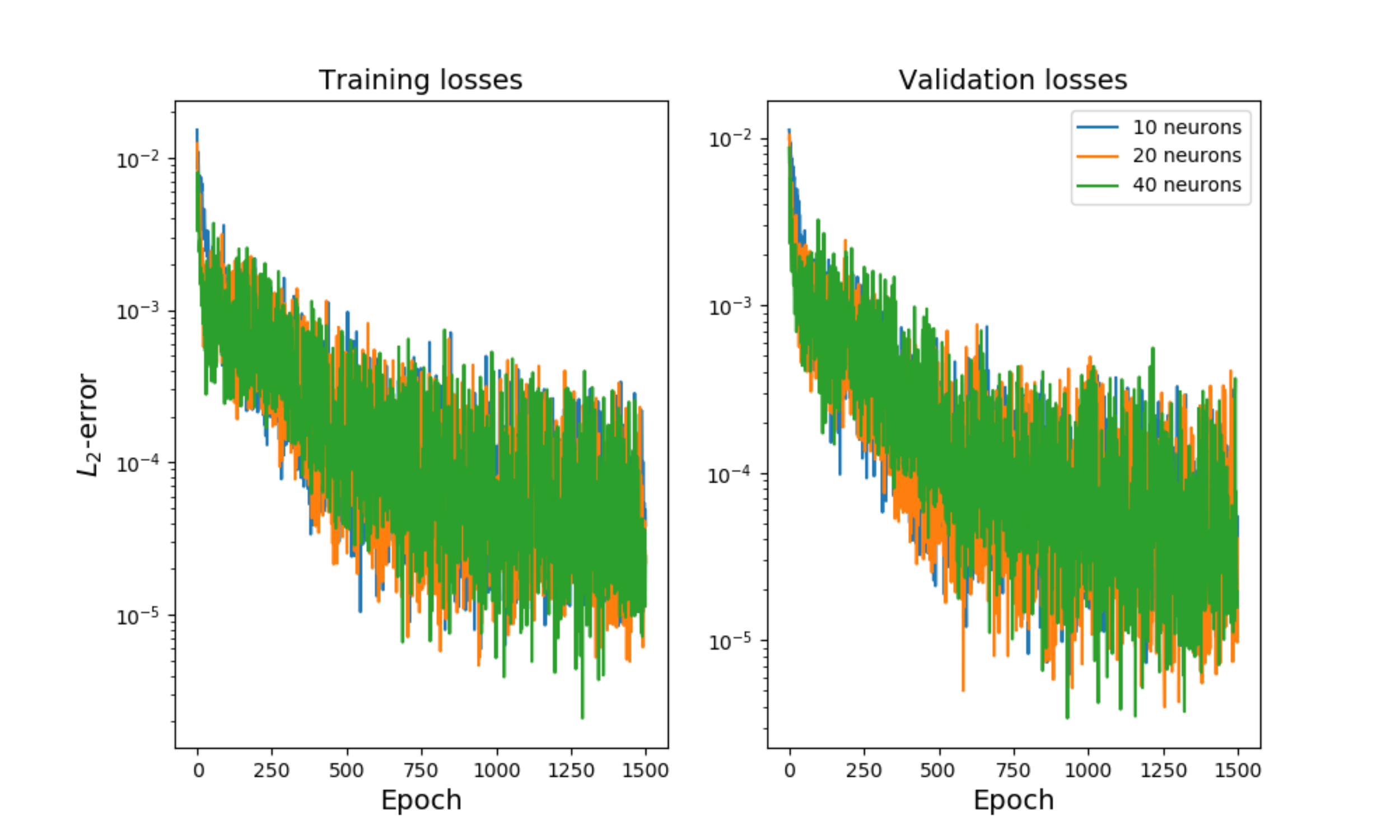}
	\caption{Ensemble training and validation losses for the NODE architecture for the advecting shock case.}
	\label{Figure8}
\end{figure}

The final time reconstructions for the true as well as the GP, LSTM and NODE time-series predictions are shown in Figure \ref{Figure4}. One can observe that at this severely truncated state, the discontinuity is not completely resolved. The GP reconstructions show the manifestation of large oscillations (now in the physical domain) whereas NODE and LSTM are able to recover the true solution well. Figures \ref{Figure5} and \ref{Figure6} show a validation of our learning in an ensemble sense, where multiple architectures (with slight differences in the hidden layer neurons) are able to recover similar trends in accuracy as examined through final time reconstruction ability. This reinforces our assumption that an implicit closure is being learned through time-series trends in a statistical manner. The corresponding training losses for the LSTM and NODE architectures are shown in Figures \ref{Figure7} and \ref{Figure8} where it can be seen similar learning trends are obtained with slight variations in the number of trainable parameters. 

\subsection{Burgers' turbulence}

\begin{figure}
	\centering
	\includegraphics[width=\textwidth]{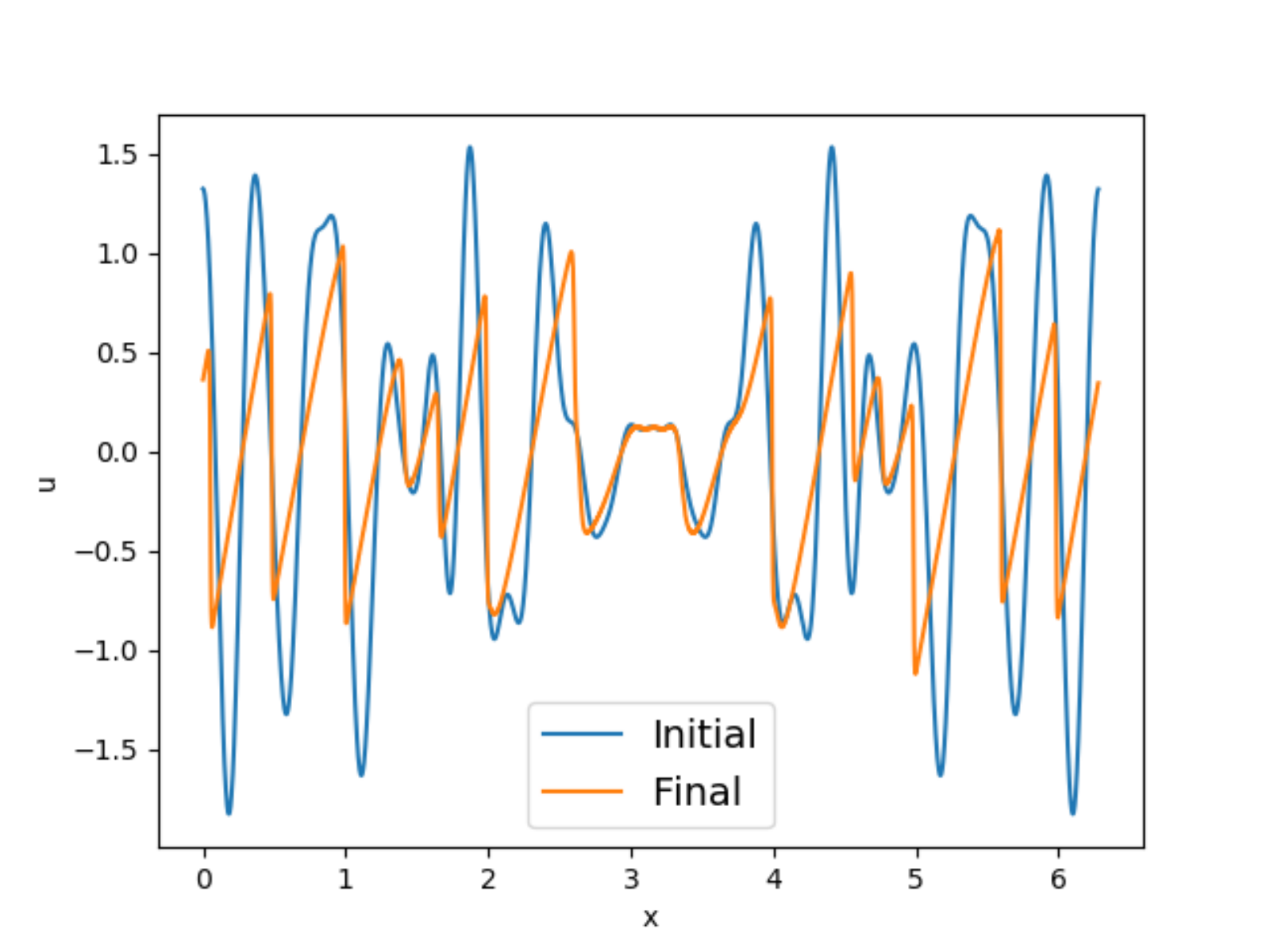}
	\caption{Initial and final conditions for the \emph{Burgulence} case showing multiple standing discontinuities decaying in strength over time.}
	\label{Figure9}
\end{figure}

Our next test case is given by the challenging Burgers' turbulence or \emph{Burgulence} test case which leads to multiscale behavior in wavenumber space. Our problem domain is given by a length, $L=2 \pi$, and the initial condition is specified by an initial energy spectrum (in wavenumber space) given by
\begin{linenomath*}
\begin{align}
E(k)=A k^{4} \exp \left(-\left(k / k_{0}\right)^{2}\right),
\end{align}
\end{linenomath*}
where $k$ is the wavenumber and $k_0 = 10$ is the parameter at which the peak value of the energy spectrum is obtained. The constant $A$ is set to the following value
\begin{linenomath*}
\begin{align}
A=\frac{2 k_{0}^{-5}}{3 \sqrt{\pi}},
\end{align}
\end{linenomath*}
in order to ensure a total energy of $\int E(k) d k=1 / 2$ at the initial condition. The initial velocity magnitudes can be expressed in wavenumber space by the following relation with our previously defined spectrum,
\begin{linenomath*}
\begin{align}
\hat{u}(k)=\sqrt{2 E(k)} \exp (i 2 \pi \Psi(k)),
\end{align}
\end{linenomath*}
where $\Psi(k)$ is a uniform random number generated between 0 and 1 at each wavenumber. Note that this distribution is constrained by $\Psi(k) = -\Psi(k)$ to ensure that a real initial condition in physical space is obtained. For the purpose of assessment, we use energy spectra given by
\begin{linenomath*}
\begin{align}
E(k, t)=\frac{1}{2}|\hat{u}(k, t)|^{2}.
\end{align}
\end{linenomath*}
The aforementioned initial conditions are solved for the viscous Burgers equation in wavenumber space by a Runge-Kutta Crank-Nicolson scheme as described in \cite{san2013stationary}. Note that our $\nu$ is chosen to be $2.5\times10^{-3}$ to ensure that sharp discontinuities emerge from the smooth initial condition. Our NODE and LSTM hyperparameters are identical to the previous test case. Our investigations here are performed for the initial condition (and its corresponding time evolution) as shown in Figure \ref{Figure9}. It may be observed that there is a considerable multiscale element to the nature of the solution - which makes this a challenging problem for POD-ROM techniques. 

\begin{figure}
	\centering
	\includegraphics[width=\textwidth]{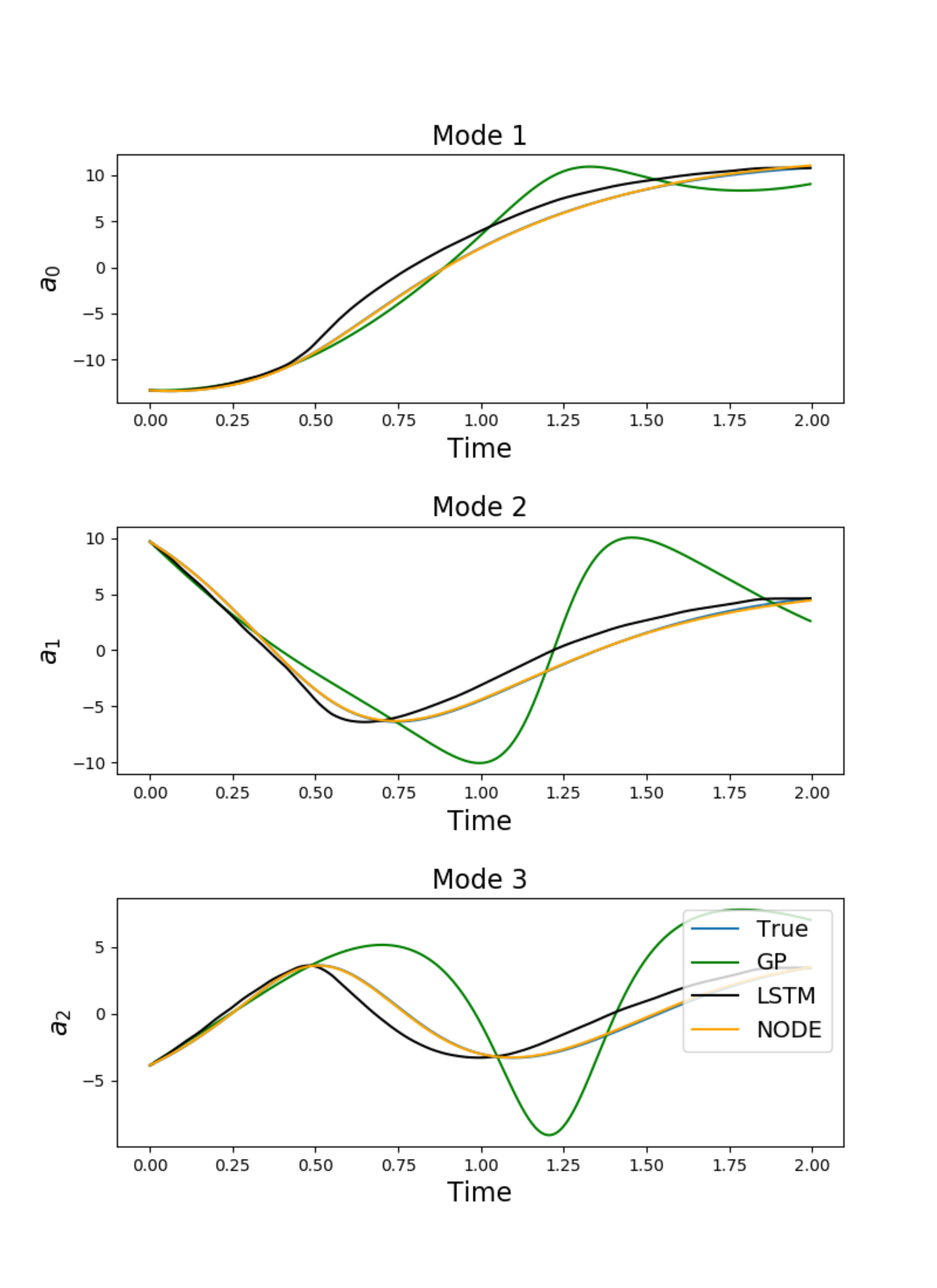}
	\caption{POD-space coefficient evolution for the \emph{Burgulence} case.}
	\label{Figure10}
\end{figure}

\begin{figure}
	\centering
	\subfigure[Spectra]{\includegraphics[width=0.48\textwidth]{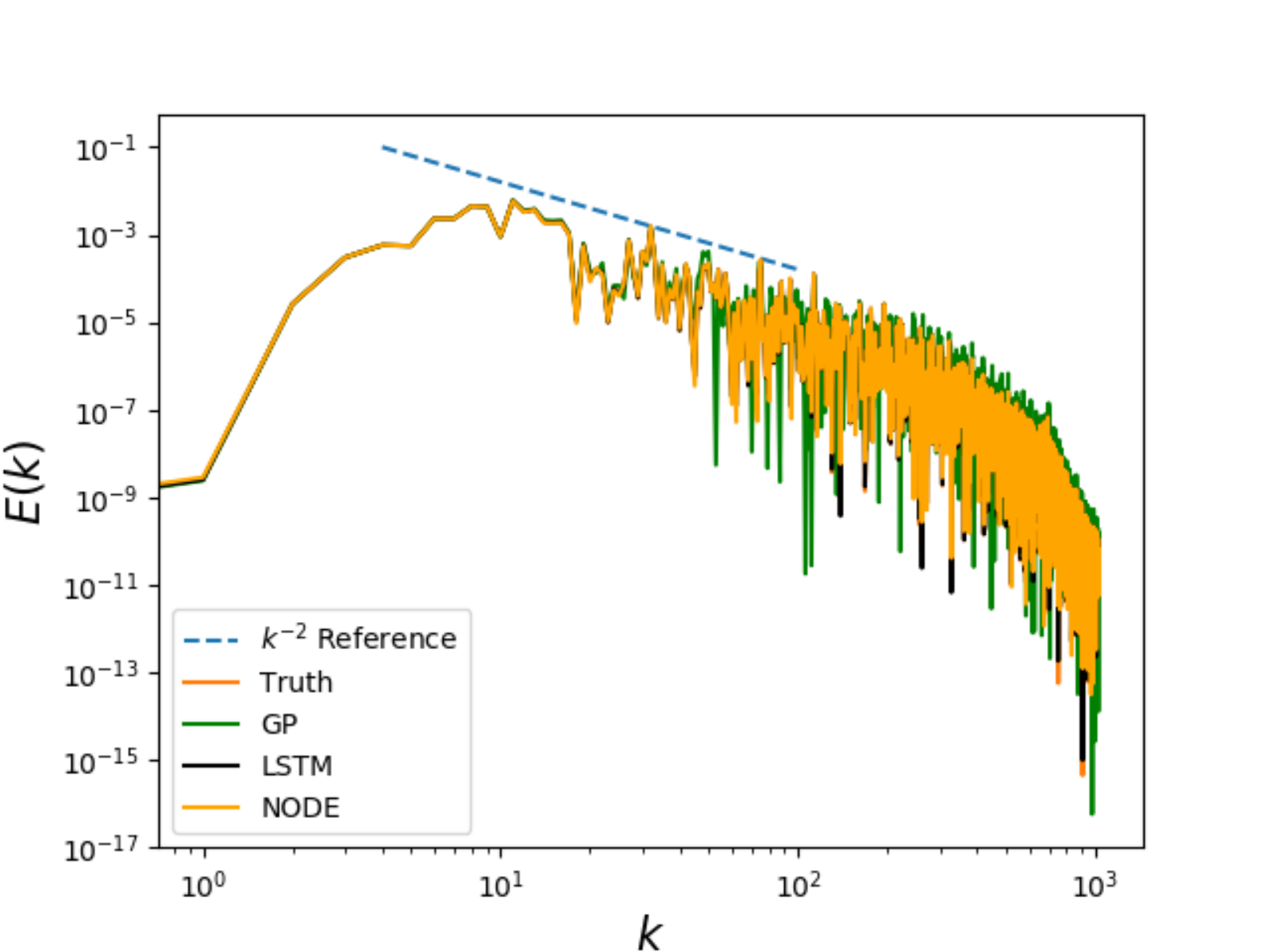}}
	\subfigure[Residual]{\includegraphics[width=0.48\textwidth]{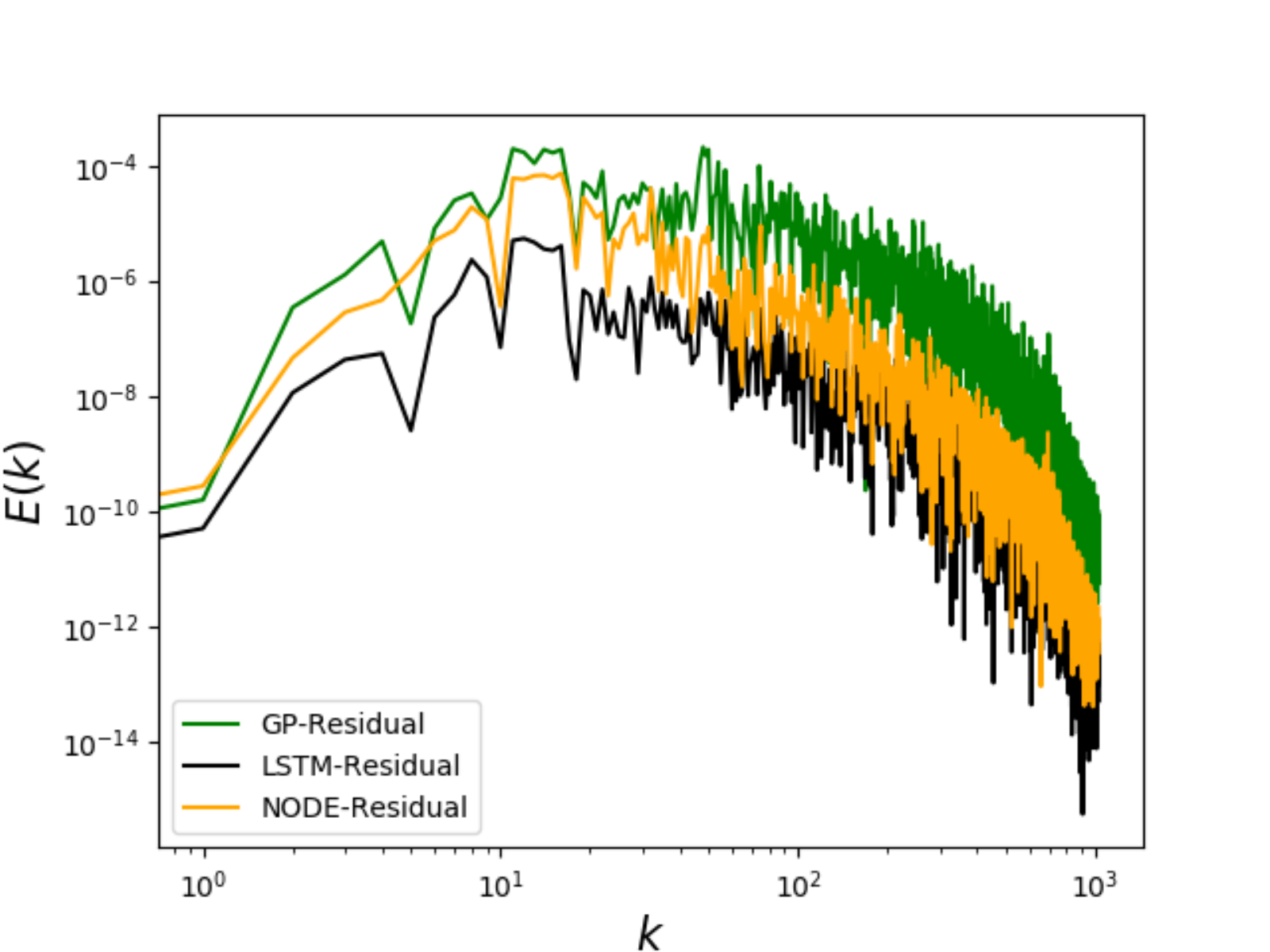}}
	\caption{Kinetic-energy spectra predictions (left) and their residuals (right) as predicted by NODE and LSTM.}
	\label{Figure11}
\end{figure}

\begin{figure}
	\centering
	\subfigure[Zoomed out]{\includegraphics[width=0.48\textwidth]{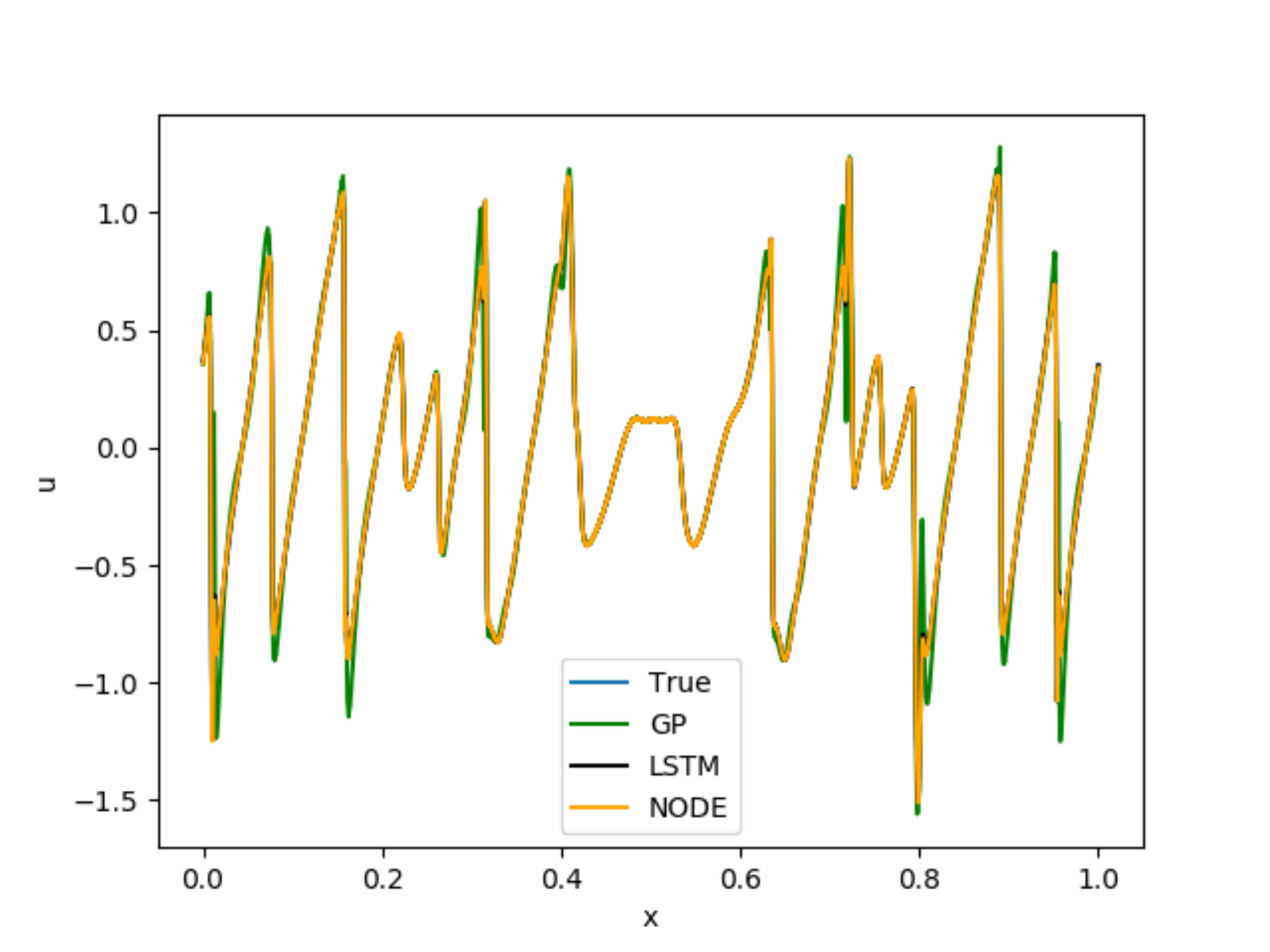}}
	\subfigure[Zoomed in]{\includegraphics[width=0.48\textwidth]{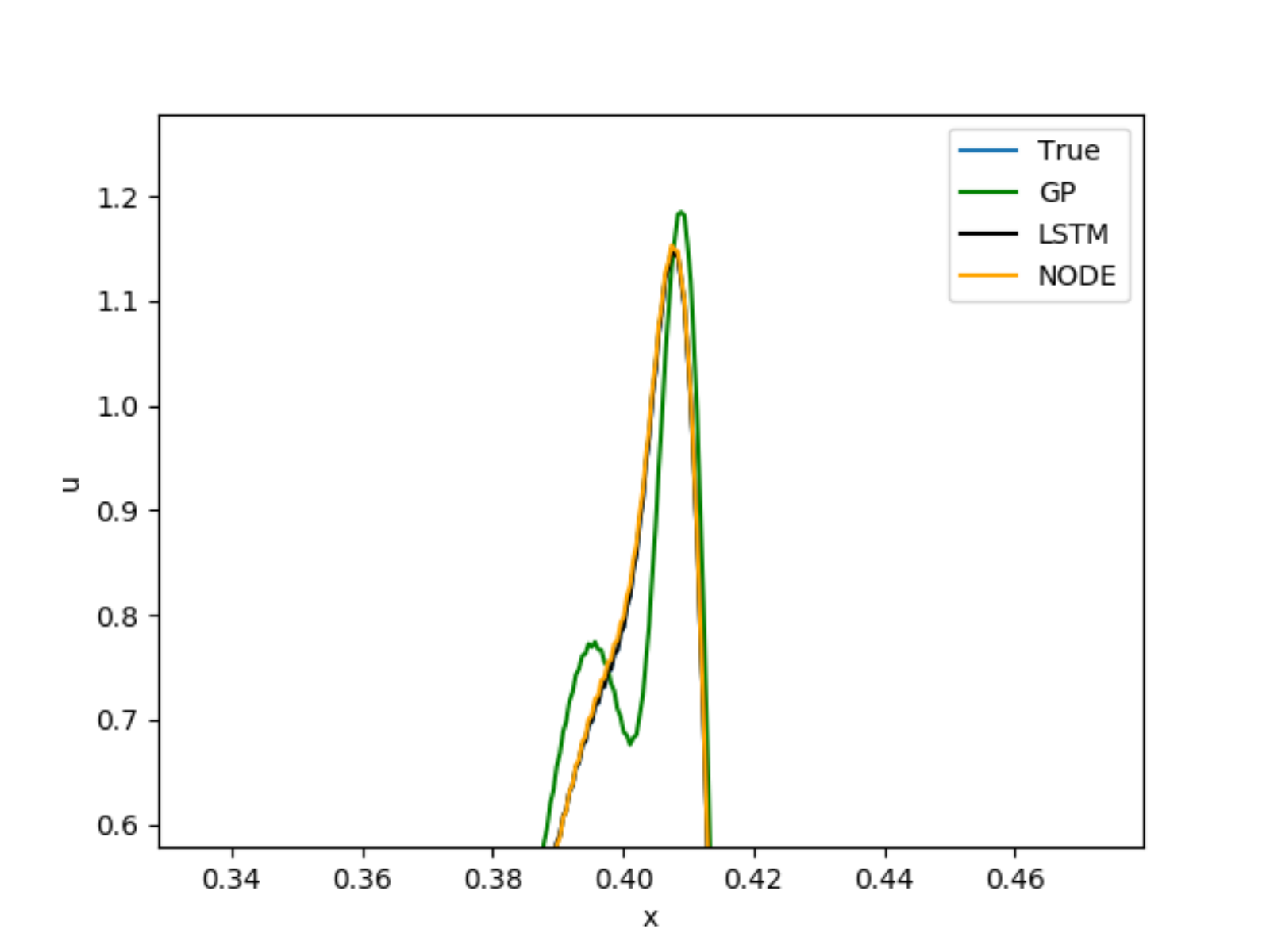}}
	\caption{Field reconstruction abilities for the NODE and LSTM frameworks showing superior performance as compared to GP.}
	\label{Figure12}
\end{figure}

Figure \ref{Figure10} shows reduced-space time-series evolutions of the three retained modal coefficients for the frameworks we are comparing. It can be observed that the LSTM and NODE techniques are successful in coefficient evolution stabilization in comparison to GP although the LSTM can be seen to add an element of phase error. The NODE, however, captures latent-space trends exceptionally well. The performance of these time-series learning models is further assessed by their reconstruction in physical space as shown in Figure \ref{Figure11} where it can be seen that the LSTM and NODE perform well in preventing spurious oscillations near discontinuities as exhibited by an unclosed GP evolution. A further validation of this hypothesis is observed in Figure \ref{Figure12} where kinetic energy spectra in wavenumber space show that the high residuals of the GP method are controlled effectively by the LSTM and NODE deployments. The LSTM is seen to result in slightly higher residuals for this particular test case and choice of hyperparameters and optimizer.

\subsection{Improving performance through hyperparameter search}

While the results obtained in the previous sections indicate an acceptable choice for hyperparameters, we utilize Deephyper \cite{balaprakash2018deephyper} to improve the test performance of our frameworks. This is motivated by the comparitively poorer performance of the LSTM in the \emph{Burgulence} experiment. Deephyper relies on an asynchronous model based search (i.e., a dynamically updated surrogate model $\mathcal{S}$ which is inexpensive to evaluate) for obtaining hyperparameters with the lowest validation losses. To ensure an expressive surrogate model which is still computationally tractable, we utilize random forests (RF). This results in a superior search algorithm as compared to both a random-search as well as a genetic algorithm based search. We note that RF also gives us the ability to handle discrete and non-ordinal parameters directly without the need for any encoding. Deephyper is configured for searching the hyperparameter space using a standard Bayesian optimization framework. In other words, for each sampled configuration $s$, $\mathcal{S}$ predicts a mean value for the validation loss $\mu(s)$ and standard deviation $\sigma(s)$. This information is utilized recursively to improve the approximation to the loss surface as predicted by $\mathcal{S}$. In terms of exploring the hyperparameter search space, evaluation points with small values of $\mu(s)$ indicate that $s$ can potentially result in the reduction of validation error subject to the accuracy of $\mathcal{S}$. Evaluation of points with large values of $\sigma(s)$ improves $\mathcal{S}$ since these locations are areas where $\mathcal{S}$ is least confident about the approximation surface. The choice for the selection of a configuration $s$ is utilized by minimizing an acquisition function given by
\begin{align}
\mathcal{A}(s)=\mu(s) - \lambda * \sigma(s)
\end{align}
where $\lambda = 1.96$ for encouraging exploration.

\begin{figure}
	\centering
	\includegraphics[width=\textwidth]{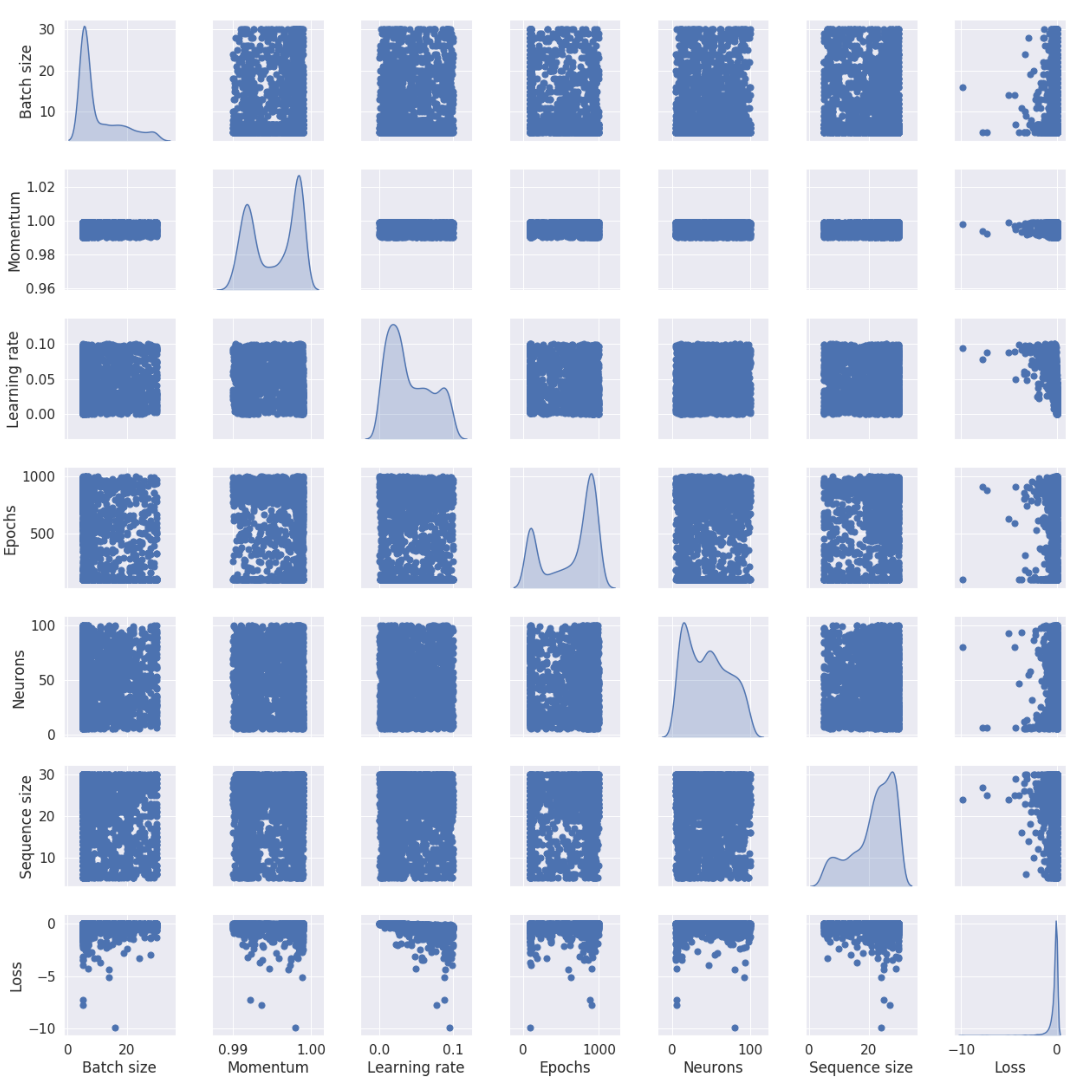}
	\caption{Pairwise dependency plots for LSTM hyperparameter search using Deephyper. Diagonal entries show distributions of configurations sampled. Note that loss is encoded as negative since the hyperparameter search is based on objective function maximization.}
	\label{Figure13}
\end{figure}

\begin{figure}
	\centering
	\includegraphics[width=\textwidth]{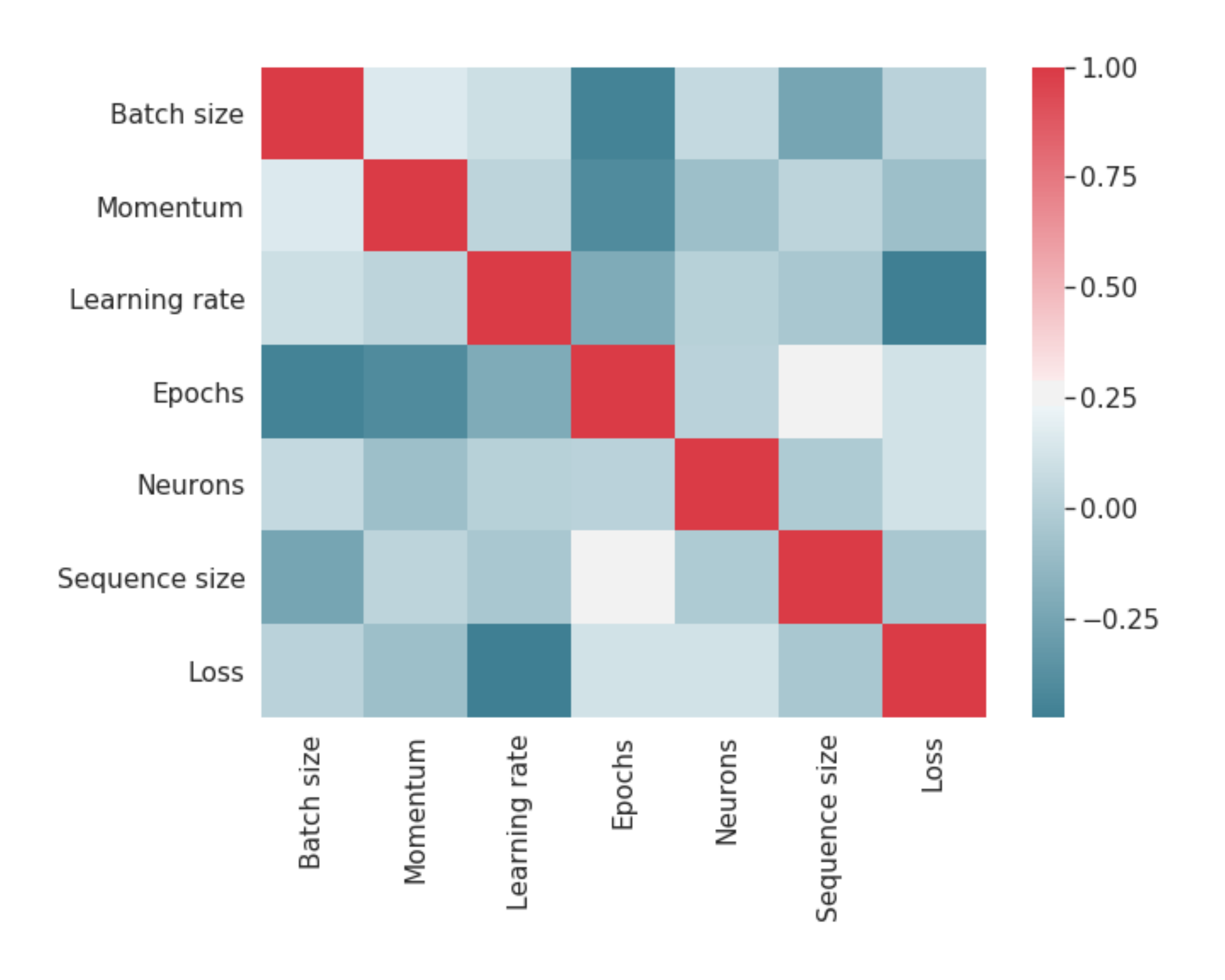}
	\caption{Pairwise LSTM hyperparameter correlations for the Burgers turbulence case.}
	\label{Figure14}
\end{figure}

Deephyper requires a range specification for real variables and a list of possible choices for discrete hyperparameters. Table \ref{Table1} outlines the range of hyperparameters for the LSTM architecture utilized for the Burgers' turbulence test case as well as the optimal hyperparameters obtained. A summary of the distribution of sampled hyperparameters and pairwise dependencies is also shown in Figure \ref{Figure13}. Note that loss is encoded as negative since the hyperparameter search is based on objective function maximization. Hyperparameter correlations are summarized in Figure \ref{Figure14} where it is observed that most hyperparameters are weakly correlated with each other. However, it must be noted that these results are problem specific. In total, 2151 hyperparameter combinations were evaluated during the process of this search.

\begin{table}[]
\begin{tabular}{|c|l|c|c|c|}
\hline
Hyperparameter & Type    & Starting value & Ending value & Optimal \\ \hline
Sequence size  & Integer & 5              & 30           & 30       \\ \hline
Neurons        & Integer & 5              & 100          & 73      \\ \hline
Learning rate  & Real    & 0.0001         & 0.1          & 0.0005 \\ \hline
Momentum       & Real    & 0.99           & 0.999        & 0.9988  \\ \hline
Epochs         & Integer & 100            & 1000         & 317     \\ \hline
Batch Size     & Integer & 5              & 30           & 8      \\ \hline
\end{tabular}
\caption{Search range for LSTM hyperparameters and their optimal values deployed for the Burgers' turbulence test case.}
\label{Table1}
\end{table}

\begin{figure}
	\centering
	\includegraphics[width=\textwidth]{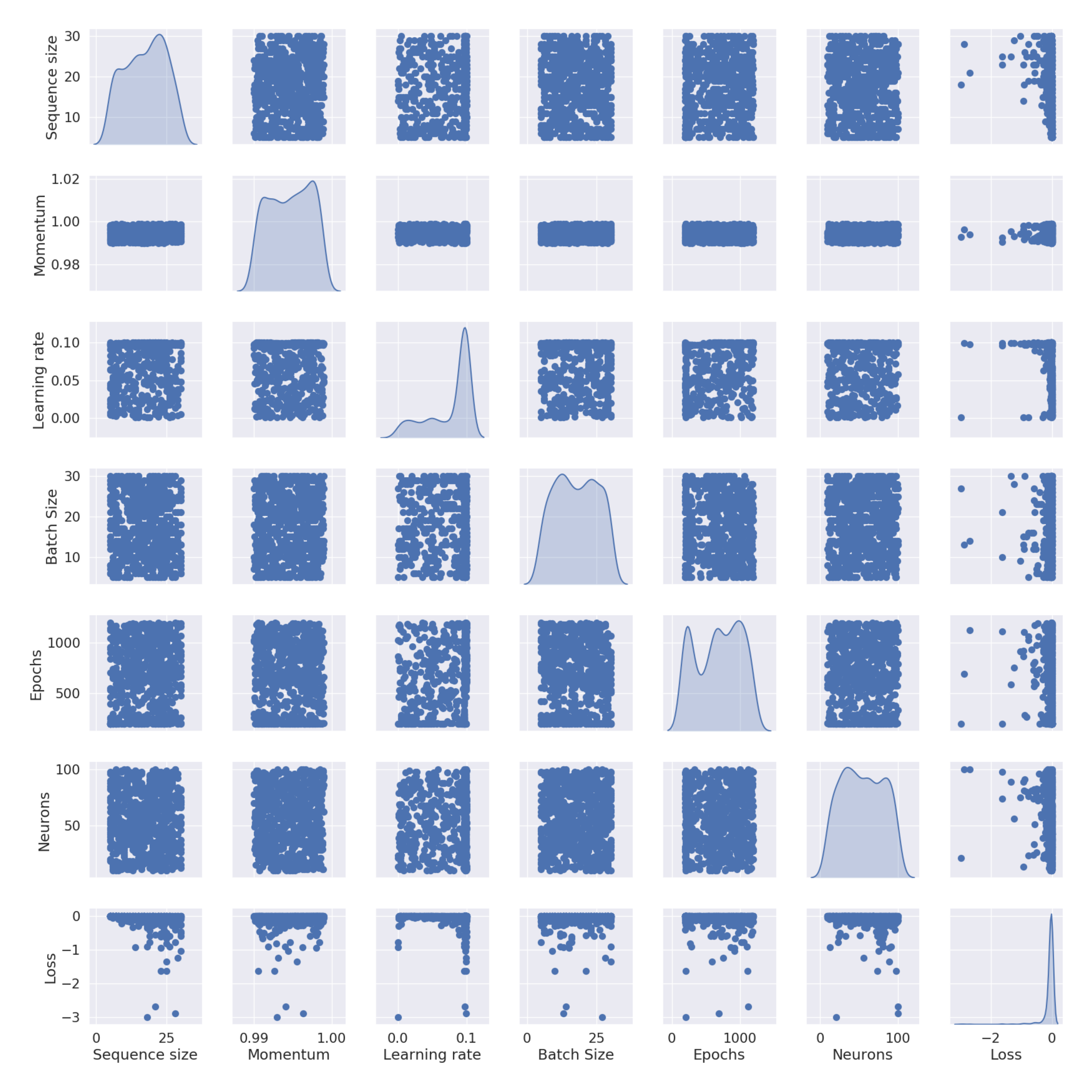}
	\caption{Pairwise dependency plots for NODE hyperparameter search using Deephyper. Diagonal entries show distributions of configurations sampled. Note that loss is encoded as negative since the hyperparameter search is based on objective function maximization.}
	\label{Figure15}
\end{figure}

\begin{figure}
	\centering
	\includegraphics[width=\textwidth]{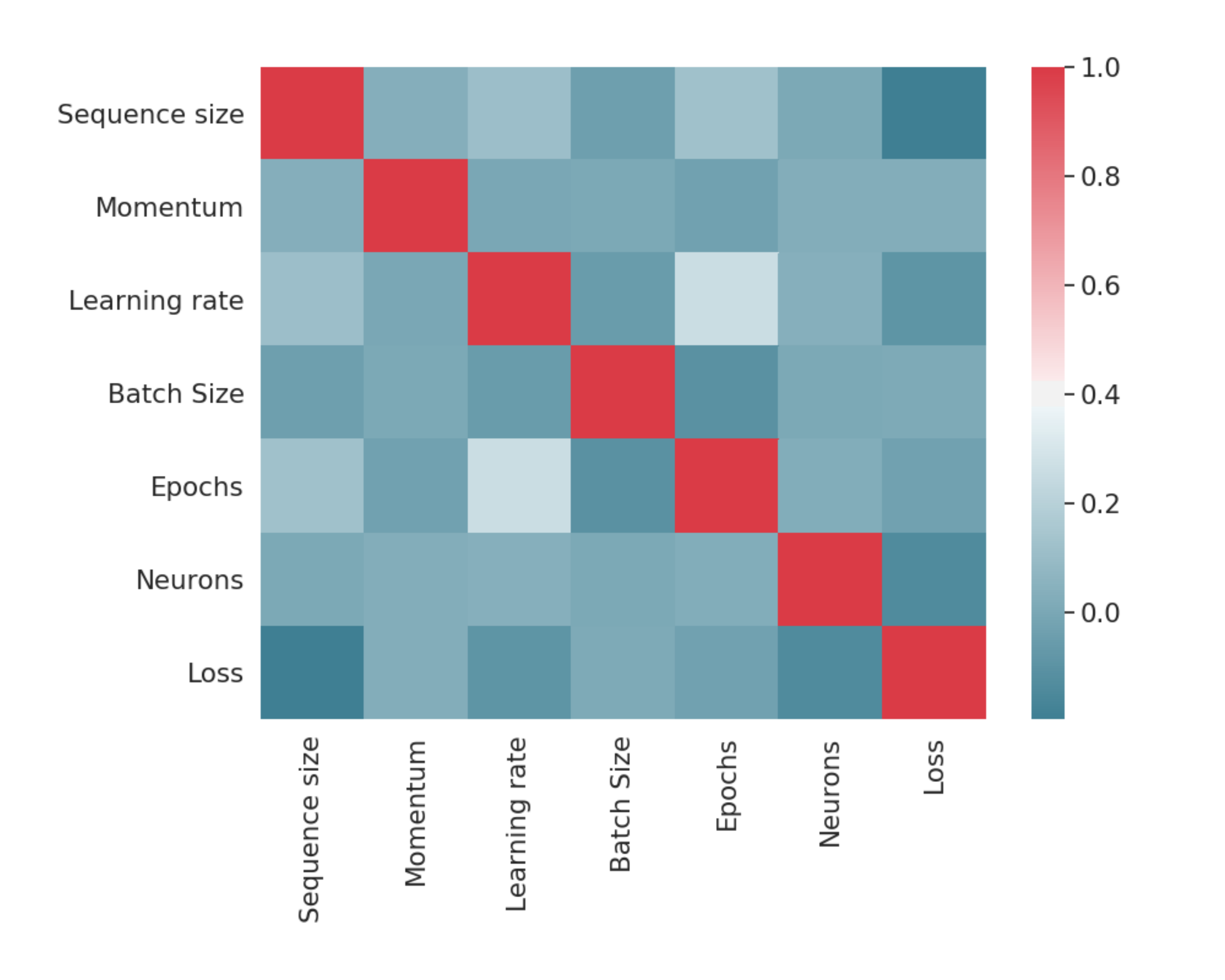}
	\caption{Pairwise NODE hyperparameter correlations for the Burgers turbulence case.}
	\label{Figure16}
\end{figure}

For the purpose of comparison, we also show results from a similar hyperparameter search experiment for the NODE but for the advecting shock experiment. The optimal parameters and ranges of this search are shown in Table \ref{Table2}. A summary of the distribution of sampled hyperparameters and pairwise dependencies is also shown in Figure \ref{Figure15}. Correlation plots between hyperparameters are shown in Figure \ref{Figure16}. In total, 734 hyperparameter combinations were evaluated during the process of this search.

\begin{table}[]
\begin{tabular}{|c|l|c|c|c|}
\hline
Hyperparameter & Type    & Starting value & Ending value & Optimal \\ \hline
Sequence size  & Integer & 5              & 30           & 5       \\ \hline
Neurons        & Integer & 10             & 100          & 82      \\ \hline
Learning rate  & Real    & 0.0001         & 0.1          & 0.0074 \\ \hline
Momentum       & Real    & 0.99           & 0.999        & 0.9983  \\ \hline
Epochs         & Integer & 200            & 1200         & 546     \\ \hline
Batch Size     & Integer & 5              & 30           & 21      \\ \hline
\end{tabular}
\caption{Search range for NODE hyperparameters and their optimal values deployed for the Burgers' turbulence test case.}
\label{Table2}
\end{table}

\begin{figure}
	\centering
	\includegraphics[width=\textwidth]{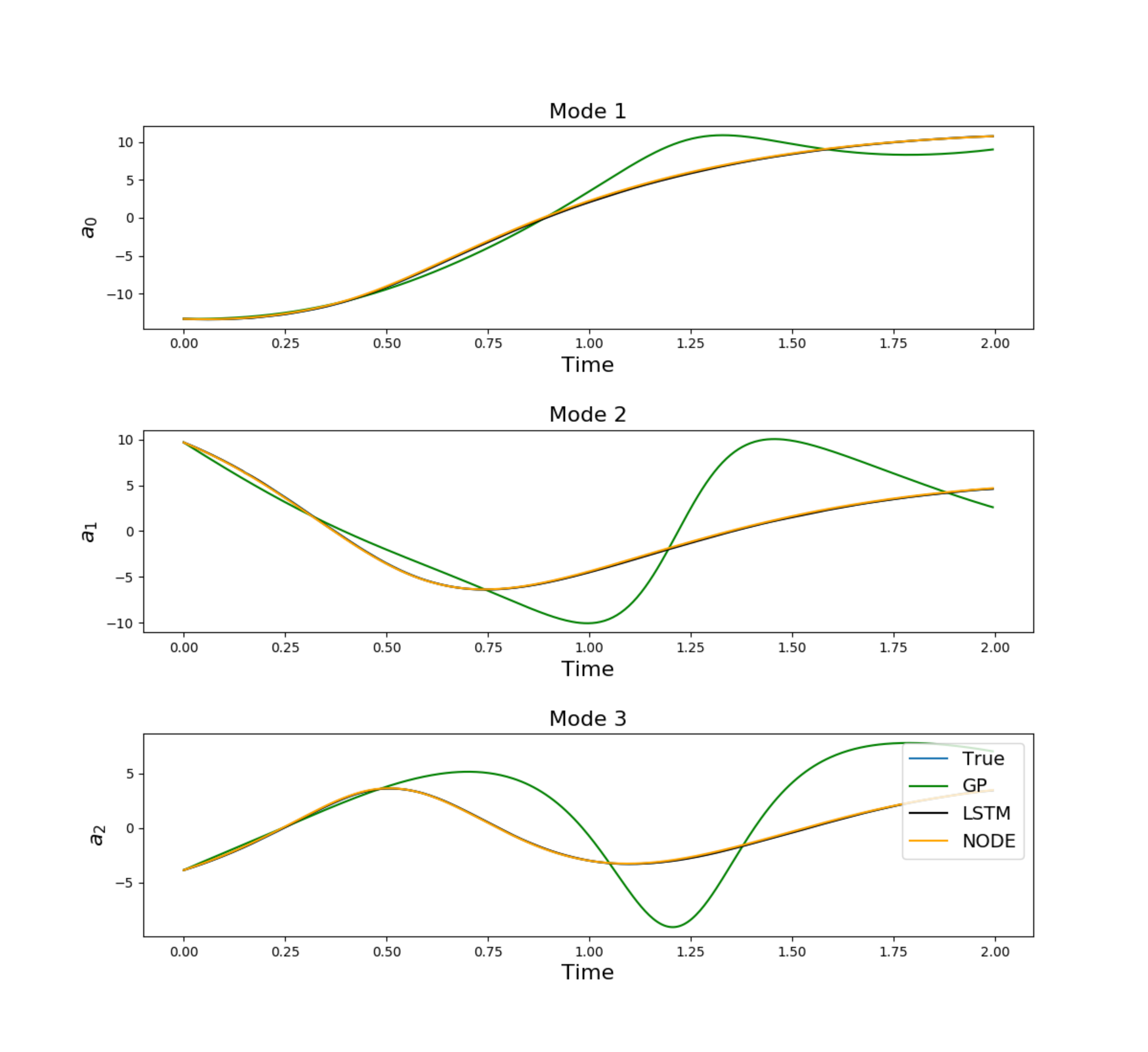}
	\caption{POD-space coefficient evolution for the \emph{Burgulence} case with improved hyperparameter choices. The LSTM performance is significantly improved.}
	\label{Figure17}
\end{figure}

\begin{figure}
	\centering
	\subfigure[Spectra]{\includegraphics[width=0.48\textwidth]{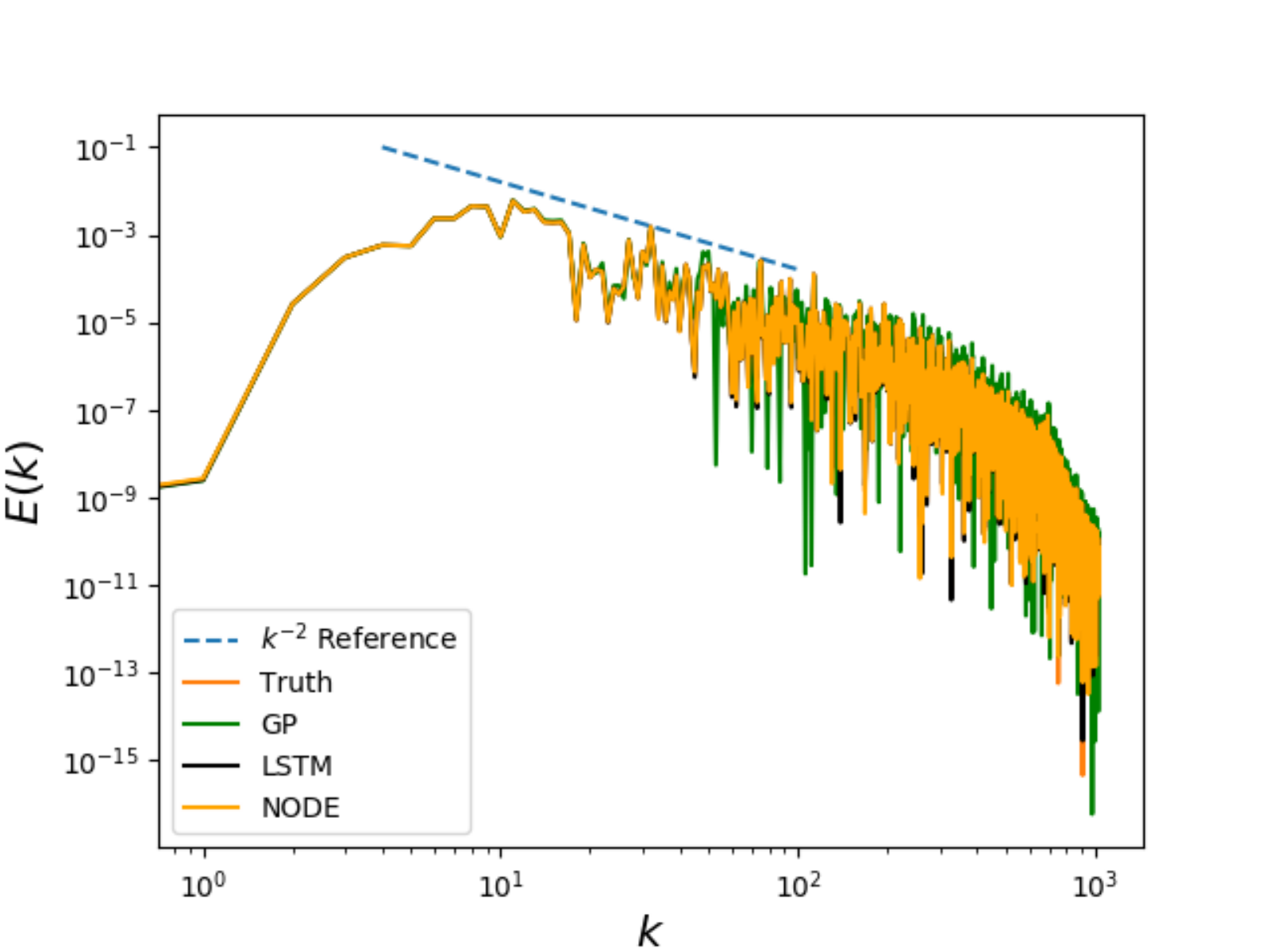}}
	\subfigure[Residual]{\includegraphics[width=0.48\textwidth]{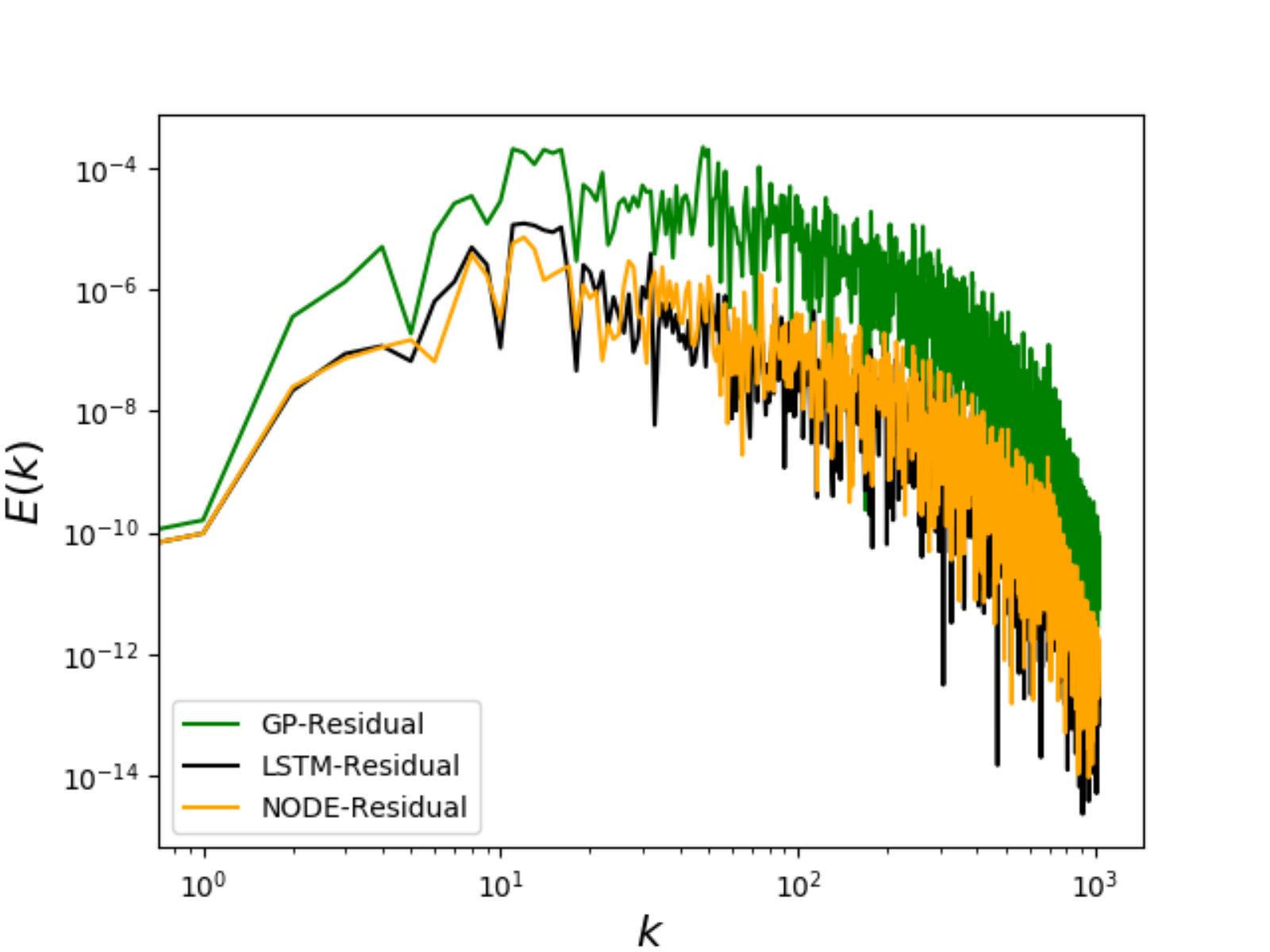}}
	\caption{Kinetic-energy spectra predictions (left) and their residuals (right) as predicted by NODE and LSTM deployed with optimal hyperparameters.}
	\label{Figure18}
\end{figure}

Finally we deploy the optimal hyperparameter configuration for an \emph{a posteriori} assessment with results as observed in Figure \ref{Figure17}. It is observed that an improved performance has been obtained using the LSTM which now matches NODE and true observations. In addition, an analysis of the spectra and residuals in Figure \ref{Figure18} confirms the superior performance as well. Deephyper has successfully led to an improved LSTM architecture.

\section{Discussion and conclusions}

In this article, we have investigated the utility of using LSTMs and NODEs as non-intrusive learning models for the projections of nonlinear partial differential equations to a latent-space spanned by severely truncated POD modes. We note that the choice of the POD modes (which form a linear subspace) also ensures the applicability of the symmetries of the PDE-governed solution on the machine-learned predictions.

Our ideas are tested on two test cases governed by the viscous Burgers' equation with the first exhibiting an advecting shock and the second displaying a multiscale nature in full-order space. Both LSTM and NODE formulations are seen to learn the transient nature of our systems in the reduced space since they exploit the sequential nature of data and end up providing an implicit closure. In the second case, we also utilize Deephyper, a scaleable Bayesian optimization package for an improved hyperparameter configuration choice in order to obtain superior performance for the LSTM. The non-i.i.d assumption of the data and associated learning allows for the embedding of a memory effect which provides for accurate coarse-grained evolution of modal coefficients in a manner similar to the Mori-Zwanzig formalism. Our assessments reveal that the machine learning techniques studied here are able to provide stable evolutions of the modal coefficients in comparison to their intrusive and unclosed counterpart (i.e., GP).

We conclude by noting that ROM developments which incorporate history explicitly (such as in the LSTM) or implicitly (such as through a NODE) represent an attractive avenue for exploration for efficient reduced basis dynamics learning of systems which are advection-dominated.

\section{Data availability}

All the relevant data and codes for this study shall be provided in a public repository at \texttt{https://github.com/Romit-Maulik/ML\_ROM\_Closures}.

\section{Acknowledgements}

This material is based upon work supported by the U.S. Department of Energy (DOE), Office of Science, Office of Advanced Scientific Computing Research, under Contract DE-AC02-06CH11357. This research was funded in part and used resources of the Argonne Leadership Computing Facility, which is a DOE Office of Science User Facility supported under Contract DE-AC02-06CH11357. This project was also funded by the Los Alamos National Laboratory, 2019 LDRD grant ``Machine Learning for Turbulence". This paper describes objective technical results and analysis. Any subjective views or opinions that might be expressed in the paper do not necessarily represent the views of the U.S. DOE or the United States Government.





\bibliographystyle{model1-num-names}
\bibliography{references.bib}







\end{document}